\documentclass[pra,aps,twocolumn,10pt,showpacs,nofootinbib,superscriptaddress]{revtex4-2}
\usepackage[utf8]{inputenc}
\usepackage[T1]{fontenc}
\usepackage{amsmath,amsthm,charter,amssymb,amsfonts,dsfont}
\usepackage{stmaryrd,mathrsfs,mathcmd,physics}
\usepackage{braket,bm,bbold,esint}

\usepackage{graphicx} 
\usepackage{xcolor}
\usepackage{tikz,rotating,float,qcircuit} 
\usepackage{todonotes}

\usepackage{ulem}
\usepackage[colorlinks=true, linkcolor=purple, citecolor=blue]{hyperref}

\usepackage{enumitem}

\newcommand{\ms}{\mathrm{s}} 

\begin{document}

    \title{Quantifying entanglement in quantum thermodynamics via separability constraints}

    \author{Joan Alba}
    \email{joan.pares@nbi.ku.dk}
    \affiliation{Niels Bohr Institute, University of Copenhagen, Jagtvej 155 A, DK-2200 Copenhagen, Denmark}

    \author{Laura Ares}
    \affiliation{Institute for Photonic Quantum Systems (PhoQS), Paderborn University, Warburger Stra\ss{}e 100, 33098 Paderborn, Germany}

    \author{Jan Sperling}
    \affiliation{Institute for Photonic Quantum Systems (PhoQS), Paderborn University, Warburger Stra\ss{}e 100, 33098 Paderborn, Germany}

    \author{Julien Pinske}
    \affiliation{Niels Bohr Institute, University of Copenhagen, Jagtvej 155 A, DK-2200 Copenhagen, Denmark}

    \date{\today}

    \begin{abstract}
        The role of quantum entanglement in thermodynamical systems remains elusive.
        Does entanglement result in thermodynamic advantages or does it impose fundamental limitations?
        Here, we unambiguously quantify the amount of heat and work in a quantum system that is due to the presence of entanglement.
        This is achieved by constraining the system's non-equilibrium dynamics to separable states, thereby isolating the impact entanglement has on thermodynamic effects.
        Unlike thermodynamic entanglement measures, which signify a loose connection between entanglement and thermodynamic properties, imposing a constraint constitutes an active intervention into a system---answering how much of a system's thermodynamics is caused by (not correlated with) its quantumness.
        We benchmark our theory by applying the constrained dynamics to several multipartite systems, including quantum batteries and quantum refrigerators.
    \end{abstract}
    
    \maketitle

    \begin{figure*}[t]
        \centering
        \begin{tikzpicture}
        \node at (0,2) {(a)};
        \node at (0,0) {\includegraphics[width=0.2\textwidth]{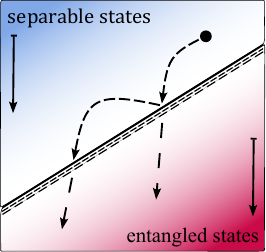}};
        \node at (-1.3,0.75) {\small{$H_\ms$}};
        \node at (1.35,-0.65) {\small{$H$}};
        \node at (5.5,2) {(b)};
        \node at (6,0) {\includegraphics[width=0.3\textwidth]{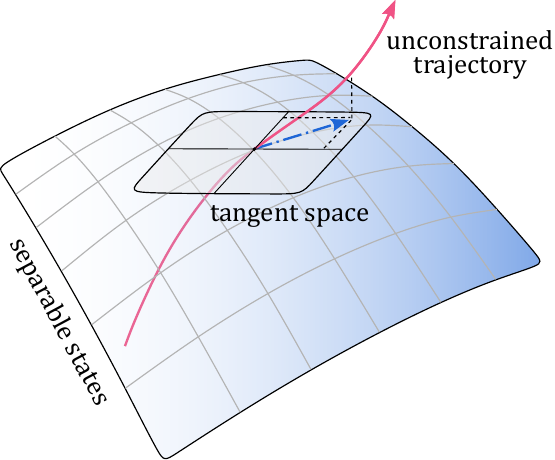}};
        \node at (12.5,2) {(c)};
        \node at (12.5, 0) {\includegraphics[width=0.28\textwidth]{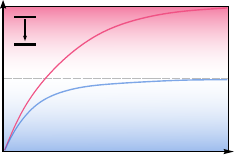}};
        \node at (12.5,-1.9) {time};
        \node at (11.1,1.3) {\small{$\ket{ee}$}};
        \node at (11.15,0.72) {\small{$\ket{\Psi^+}$}};
        \node[rotate=90] at (9.7,0) {\small{$\braket{\Psi^+|\rho|\Psi^+}$}};
        \node at (13,1) {\small{$\rho(t)$}};
        \node at (13,-0.35) {\small{$\rho_\ms(t)$}};
        \node at (15,-1.8) {\small{$t$}};
        \node at (15.05,-0.05) {$\frac{1}{2}$};
        \node at (15.05,1.5) {$1$};
        \end{tikzpicture}
        \caption{\label{fig:synopsis} (a) Trajectory governed by a constrained Hamiltonian $H_\ms$ contrasted with the freely evolving trajectory governed by the Hamiltonian $H$. 
        (b) Schematic representation of a constraint. 
        In general, the unrestricted evolution takes a state outside the manifold of separable states. 
        For constrained dynamics, at each time step, the evolving state is projected onto the tangent space of separable states. 
        (c) Correlated decay of a doubly-excited state $\ket{ee}$ into a Bell state $\ket{\Psi^+}=(\ket{ge}+\ket{eg})/\sqrt{2}$.
        While the freely evolving state $\rho$ approaches $\ket{\Psi^+}$, the constrained state $\rho_\ms$ remains separable at all times, e.g., $\braket{\Psi^+|\rho_\ms|\Psi^+}\leq 1/2$. 
        }
    \end{figure*}
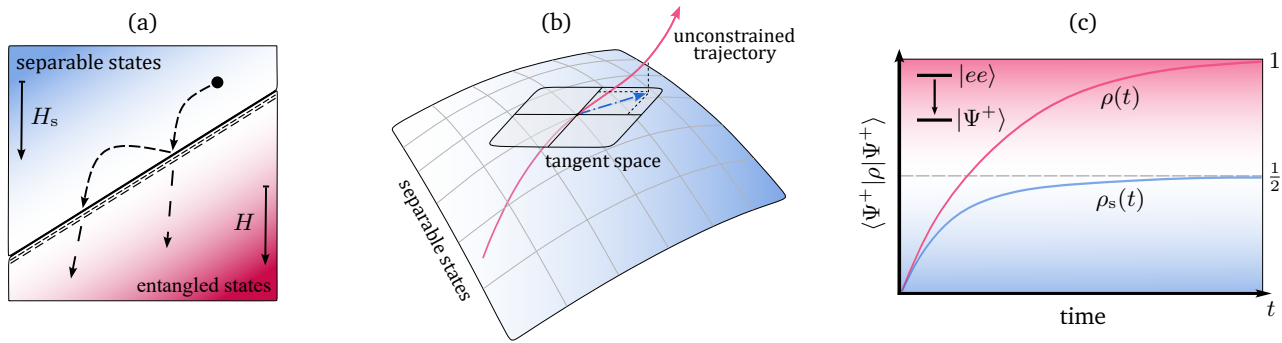

    \section{Introduction}
    \label{sec:intro}

    The often counterintuitive behavior of quantum systems emerges when exploring increasingly smaller length and energy scales.
    Quantum physics is therefore traditionally concerned with the behavior of microscopic systems, for which quantum correlations become a dominant effect. 
    Thermodynamics, by contrast, aims at describing the macroscopic world, where the collective behavior of vast numbers of particles results in emergent laws governing energy, entropy, and irreversibility.
    The ongoing miniaturization of modern technology, as well as the design and control of exceedingly complex quantum systems, motivates a unified understanding of the thermodynamical properties of quantum mechanical systems, giving rise to the emergent field of quantum thermodynamics \cite{VA16}.
    
    A primary concern in the thermodynamic treatment of quantum systems is whether quantum correlations, such as coherent superpositions \cite{A06,BC14}, quantum entanglement \cite{W89,HHH09}, and contextuality \cite{S05}, result in thermodynamic advantages or barriers in the design of quantum engines \cite{SS59,QLS07}, batteries \cite{BVM15}, and refrigerators \cite{PKG01,ME12}.
    In thermodynamical systems, entanglement has been linked to an improvement in the cooling capabilities of quantum refrigerators \cite{CPA13,BHL14,LPS10,SBLP11}, a speedup in the charging time of a quantum battery \cite{AF13,SGZ25}, as well as reversals of the thermodynamic arrow of time \cite{P08,JR10}.
    Nevertheless, the precise contribution of entanglement to a thermodynamic quantity remains elusive.
    For example, entanglement being a resource for the optimal extraction of work has been argued for \cite{FWU13,GSR22} and against \cite{HPH13}. 
    This is partly due to a lack of a genuine reference to which the non-equilibrium dynamics of an entangled quantum system can be compared.
    Existing studies are exemplary, relying on the choice of some non-entangling benchmark to which the performance of the entangling device is then compared.
    The proposed entanglement advantage can then depend significantly on the (arbitrary) choice of benchmark, making it ambiguous.
    Another approach alters the physical control parameters of the system under investigation, e.g., frequencies, decay rates, temperatures, etc., to move between parameter regimes for which the system is either entangled or separable \cite{BHL14,CPA13}. 
    However, changing the physical parameters' regime changes more than just entanglement; it can alter other properties, such as coherences and adiabaticity.
    This makes the genuine contribution that entanglement gives to the quantum advantage gained by a system hard to track.
    A genuine quantification of the contribution of entanglement to the thermodynamics of a system, e.g., heat, work, entropy, etc., requires one to compare it to a situation in which the system performs the same task without using entanglement.

    In this article, we unambiguously quantify the thermodynamic properties of a quantum system that are due to entanglement between its subsystems.
    This is achieved by imposing a constraint on a quantum state to evolve through separable states only \cite{SW17,PA24,AP24}; see Fig.~\ref{fig:synopsis} (a).
    In mechanical systems, a (holonomic) constraint is associated with a very strong force field directed towards a surface (or curve).
    In the limit of an infinitely strong force, the trajectory of a particle must remain on that surface, confining the particle's motion \cite{A10}, e.g., the stiff suspension of a pendulum and the fixed position of point masses in a rigid body [Fig.~\ref{fig:synopsis} (a)].
    Constraints extend naturally to quantum mechanical systems, with one notable example being the quantum Zeno effect \cite{MS77} by which frequent measurement of a quantum system prevents it from evolving outside a certain subspace.
    Indeed, this provides an operational meaning for dynamics constrained to separable states; that is, the constraint is realized by a strong measurement projecting onto the closest product state of a quantum trajectory at each instant in time [Fig.~\ref{fig:synopsis}~(b)].
    This highlights that the constrained evolution will generally possess different thermodynamics as repeated measurements have been identified as a source of thermodynamic work \cite{EHH17,B25}.
    
    The restriction to separable states removes entanglement as a dynamical resource while keeping all other aspects of the system, such as coherent driving, coupling to baths, and joint classical correlations, unchanged. 
    The constrained dynamics then provide a natural reference, against which we can directly compare the (unconstrained) evolution that includes entanglement; see Fig.~\ref{fig:synopsis}~(c) for an example.
    We stress that the scope of this work goes beyond the, by now, well-established strategy of using thermodynamic quantities as a type of entanglement witness \cite{AKL06,WVB08,OBL25,P94}; although, verifying the presence of entanglement is a by-product of the constrained evolution being different from the freely evolving one. 
    Our aim here is more ambitious.
    How much heat, work, etc. would a quantum system generate if it could only make use of its classical correlations?
    This is a question that cannot be resolved by referring to entanglement measures.
    This path enables a transparent analysis of how entanglement influences key performance indicators, such as heat currents, power output, and efficiency in quantum thermodynamic devices.
    
    As part of our investigation, we formulate the first and second laws of quantum thermodynamics for systems constrained to separable states. 
    Using this framework, we show that forcing a quantum battery to charge through separable states limits the work that can be extracted. 
    We further demonstrate that in a three-qubit quantum refrigerator, all cooling advantage is due to entanglement, not just part of it.
    Finally, we identify situations in which system-bath entanglement actually limits the amount of heat exchange rather than amplifying it. 
    Together, these results provide a far-reaching analysis of the role of entanglement in nonequilibrium quantum thermodynamics beyond what can be inferred from entanglement measures alone.
    
    The structure of the article is as follows. 
    Section \ref{sec:quant-therm} reviews the thermodynamics of open quantum systems. 
    In Sec.~\ref{sec:sep-evolve}, we derive equations of motion for a system constrained to separable states.
    In Sec.~\ref{sec:sep-thermo}, we formulate the first and second laws of quantum thermodynamics for constrained systems.  
    Section~\ref{sec:apply} applies our theory to the non-equilibrium dynamics of several benchmark systems.
    This includes a three-qubit quantum refrigerator, correlated dephasing in a two-qubit system, and the impact of system-bath entanglement on a qubits heat exchange. 
    Finally, Sec.~\ref{sec:fin} is reserved for a summary of the article and concluding remarks.

    \section{Thermodynamics of open quantum systems}
    \label{sec:quant-therm}

    The interplay of quantum correlations and the statistical properties of a system are manifest in the first law of thermodynamics,
    \begin{equation}
        \label{eq:1st-law}
        dE = \delta Q + \delta W,
    \end{equation}
    which relates the change in the internal energy $E$ to the (extractable) work $W$ and the heat $Q$.
    For a quantum system in the state $\rho(t)$ with Hamiltonian $H(t)$, the internal energy is identified with the quantum mechanical mean value $E=\braket{H}_{\rho}=\Tr(\rho H)$.
    Taking the time derivative of $E$ and comparing to Eq.~\eqref{eq:1st-law}, we identify 
    \begin{equation}
        \label{eq:heat-work}
        \dot{Q}=\Tr(\dot{\rho} H),\qquad \dot{W}=\mathrm{Tr}(\rho\dot{H}),
    \end{equation}
    as changes in heat and work, respectively.
    The quantity $\dot{W}$ captures the explicit time dependence of a Hamiltonian and thus accounts for the average change of the system’s energy due to, for instance, external driving and on-off interactions.
    The quantity $\dot{Q}$ describes the energy intake (and release) caused by changes in the system’s state.
    
    The most general change a state $\rho_0$ can undergo is given by a dynamical map, $\rho(t)=\Lambda_t(\rho_0)$, which accounts for coherent driving as well as the system’s contact to one or multiple thermal reservoirs $B$.
    The unitary time evolution $U(t)$ of the joint system-environment complex yields the dynamical map
    \begin{equation}
        \label{eq:dyn-map}
        \rho(t) = \Tr_{B}[U(t)(\rho_0\otimes \rho_B)U(t)^\dag],
    \end{equation}
    where the environment is initially in the state $\rho_B$.
    For a time-independent Hamiltonian, $\dot{H}=0$, the heat flow in Eq.~\eqref{eq:heat-work} can be integrated, leading to
    \begin{equation}
        \label{eq:heat}
        Q(t)=\Tr\{[\rho(t)-\rho_0]H\}.
    \end{equation}
    Since no external work is performed in this scenario, $W=0$, the accumulated heat corresponds to the change in internal energy, i.e., $Q(t)=E(t)-E(0)$.
    
    \subsection{Irreversibility and entropy production}
    
    The reduced dynamics of a quantum system are, in general, irreversible.
    To quantify the irreversible character of an evolution, we note that the relative entropy
    \begin{equation}
        \label{eq:rel-ent}
        S(\rho||\rho^\prime)=\Tr(\rho\ln\rho) - \Tr(\rho\ln\rho^\prime),
    \end{equation}
    is nonincreasing under the dynamical map~\eqref{eq:dyn-map} \cite{L75}, i.e., 
    \begin{equation}
        \label{eq:ent-ineq}
        S[\Lambda_t(\rho)||\Lambda_t(\rho^\prime)] \leq S(\rho||\rho^\prime).
    \end{equation} 
    
    If the interaction between the system and the environment is Markovian, then the state $\rho$ obeys a quantum master equation in Lindblad form \cite{L76}
    \begin{equation}
        \label{eq:LMEq}
        \begin{split}
            \frac{d\rho}{dt} & = \mathcal{L}(\rho)\\
             & = i[\rho,H] + \sum_k \mathcal{D}_k(\rho).\\
        \end{split}
    \end{equation} 
    The Hamiltonian $H$ of the system yields a unitary contribution to the overall evolution, and 
    \begin{equation}
        \mathcal{D}_k(\rho) = L_k\rho L_k^\dag - \frac{1}{2}\{L_k^\dag L_k,\rho\}
    \end{equation}
    accounts for dissipation due to jump operators $L_k$.

    For concreteness, consider a system with time-independent Hamiltonian $H$ that is coupled to a thermal bath, such that the thermal (equilibrium) state, $\gamma=e^{-\beta H}/Z$, is the dynamics' (unique) steady state, i.e., $\mathcal{L}(\gamma)=0$.
    Here, $Z=\Tr(e^{-\beta H})$ is the partition function and $\beta=1/T$ is the inverse temperature of the bath (we set $k_{\mathrm{B}}=1$). 
    For a Markovian dynamical map, $\Lambda_t=\exp(\mathcal{L}t)$, we have $\Lambda_t(\gamma)=\gamma$, for all $t$.
    It follows from Eq.~\eqref{eq:ent-ineq} that the entropy production rate is nonnegative \cite{S78,BP10},
    \begin{equation}
        \label{eq:ent-prod}
        \sigma(t) = - \frac{d}{dt}S[\rho(t)||\gamma] \geq 0.
    \end{equation}
    As $\sigma$ is nonnegative, $\rho(t)$ approaches the thermal state $\gamma$ in a monotonic manner, without any (non-Markovian) backflow of information \cite{BLP02}.
    Moreover, the inequality~\eqref{eq:ent-prod} expresses the second law of thermodynamics as
    \begin{equation}
        \label{eq:2nd-law}
        \dot{S} - \beta\dot{Q} \geq 0,
    \end{equation}
    where $S(\rho)=-\Tr(\rho\ln\rho)$ is the von Neumann entropy and $\dot{Q}=\sum_k\Tr[\mathcal{D}_k(\rho) H]$ originates from dissipation.

    \section{Dynamics constrained to separable states}
    \label{sec:sep-evolve} 
    
    The state $\ket{\psi}$ of a bipartite system $\mathcal{H}_A\otimes \mathcal{H}_B$ evolves according to the principle of stationary action, in which the action $S=\int_{0}^T L dt$ is minimized for the Lagrangian
    \begin{equation}
        \label{eq:L-Schroed}
        L(\psi,\dot{\psi})= \frac{i}{2}\big(\braket{\psi|\dot{\psi}} - \braket{\dot{\psi}|\psi}\big) - \braket{\psi|H|\psi}.
    \end{equation}
    Applying variational calculus, a stationary action implies that $L$ obeys the Euler-Lagrange equation
    \begin{equation}
        \label{eq:ELEq}
        \frac{d}{dt}\frac{\partial L}{\partial \bra{\dot{\psi}}} - \frac{\partial L}{\partial \bra{\psi}} = 0.
    \end{equation}
    Inserting Eq.~\eqref{eq:L-Schroed} into Eq.~\eqref{eq:ELEq} verifies that $\ket{\psi}$ obeys Schrödinger's equation, $i\ket{\dot{\psi}}=H\ket{\psi}$, with $\hbar=1$. 

    To constrain the evolution to separable states, we impose that $\ket{\psi(t)}=\ket{\psi_A(t)}\otimes\ket{\psi_B(t)}=\ket{\psi_A(t)\psi_B(t)}$ remains a product state at all times. 
    The local states $\ket{\psi_A}$ and $\ket{\psi_B}$ play the role of generalized coordinates satisfying the constraint.
    The Euler-Lagrange equation for subsystem $A$ is
    \begin{equation}
        \frac{d}{dt}\frac{\partial L}{\partial \bra{\dot{\psi}_A}} - \frac{\partial L}{\partial \bra{\psi_A}} = 0,
    \end{equation}
    and---in analogy---for subsystem $B$
    \begin{equation}
        \frac{d}{dt}\frac{\partial L}{\partial \bra{\dot{\psi}_B}} - \frac{\partial L}{\partial \bra{\psi_B}} = 0.
    \end{equation}
    Inserting the Lagrangian~\eqref{eq:L-Schroed} into the above equations and using the constraint $\ket{\psi}=\ket{\psi_A\psi_B}$ yields the separability Schrödinger equations \cite{SW17}
    \begin{equation}
        \label{eq:SSE}
        i\frac{d}{dt}\ket{\psi_A}=(H)_A\ket{\psi_A},\quad i\frac{d}{dt}\ket{\psi_B}=(H)_B\ket{\psi_B}, 
    \end{equation}
    where 
    \begin{equation}
        \label{eq:red-ops}
        (H)_A=\frac{\braket{\psi_B|H|\psi_B}}{\braket{\psi_B|\psi_B}},\qquad (H)_B=\frac{\braket{\psi_A|H|\psi_A}}{\braket{\psi_A|\psi_A}}
    \end{equation}
    are reduced operators acting on subsystems $A$ and $B$, respectively.
    These are coupled nonlinear differential equations, which describe the evolution of a composite system restricted to separable states $\ket{\psi(t)}=\ket{\psi_A(t)\psi_B(t)}$. 

    By contrasting the two evolutions---one constrained to separable states, the other free to explore the full state space--- we obtain an unambiguous measure of the impact entanglement has on the dynamics.
    Let us stress that we do not simply take an entangling $H$ and remove the interaction between the subsystems $A$ and $B$.
    Rather, we force the interaction of the two subsystems to be mediated without using entanglement, e.g., using local coherences and shared classical correlations. 

    Finally, note that for a local Hamiltonian, $H=H_A\otimes \mathbb{1} + \mathbb{1}\otimes H_B$, the operators in Eq.~\eqref{eq:red-ops} are just $(H)_A=H_A$ and $(H)_B=H_B$.
    Then, the constrained dynamics in Eq.~\eqref{eq:SSE} coincide with the conventional Schrödinger equation \cite{SW17}.

    \subsection{Generalization to multipartite systems}

    The separability Schrödinger equation applies to multipartite systems as well \cite{SW17,SW20}.
    For an $N$-partite product state $\ket{\psi_1\dots\psi_N}$, the constrained dynamics are subject to the coupled equations $i\ket{\dot{\psi}_k}=(H)_k\ket{\psi_k}$, with 
    \begin{equation}
    \label{eq:sep-op}
        (H)_k=\frac{(\otimes_{j\neq k}\bra{\psi_j})H(\otimes_{j\neq k}\ket{\psi_j})}{\prod_{j\neq k}\braket{\psi_j|\psi_j}}
    \end{equation}
    being the reduced Hamiltonian of the $k$th subsystem, and $j=1,\dots, N$.
    With this notation at hand, we can introduce the constrained Hamiltonian 
    \begin{equation}
        \label{eq:constrained-H}
        H_{\ms}=\sum_{k=1}^N \mathbb{1}^{\otimes (k-1)}\otimes (H)_k\otimes \mathbb{1}^{\otimes (N - k)},
    \end{equation}
    which allows us to formulate the separable dynamics as a von Neumann-type equation,
    \begin{equation}
        \label{eq:sep-VN}
        \frac{d\rho_{\ms}}{dt}=i[\rho_{\ms}, H_{\ms}],
    \end{equation}
    where $\rho_{\ms}=\ket{\psi_1\dots\psi_N}\bra{\psi_1\dots\psi_N}$ is the constrained state.

    \begin{figure}[t]
        \centering
        \begin{tikzpicture}
        \node at (1.5,0) {\includegraphics[width=0.25\textwidth]{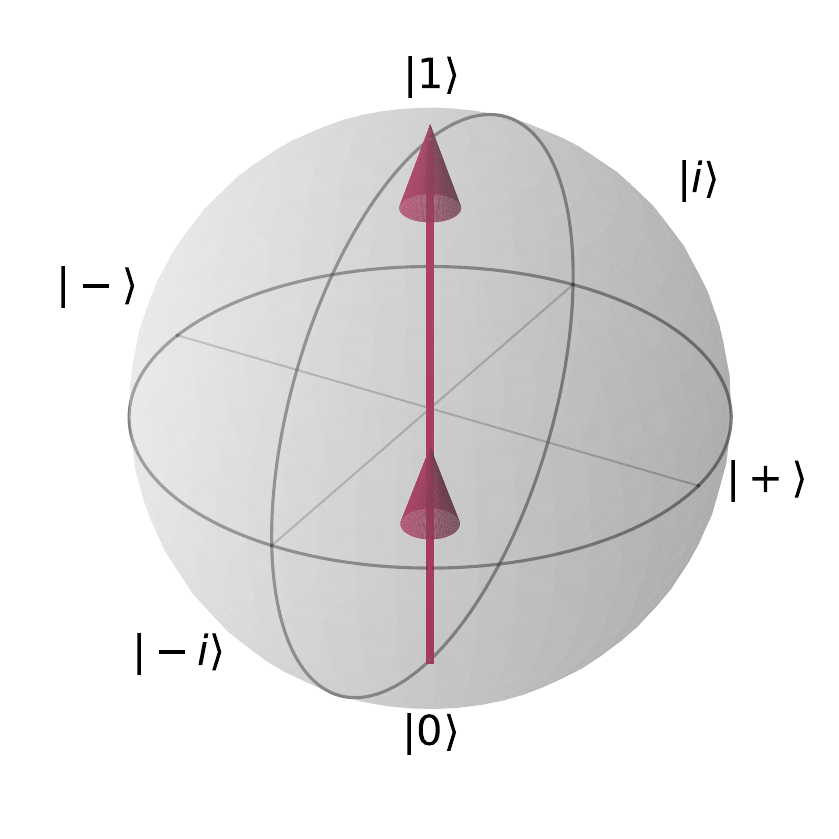}};
        \node at (0,1.8) {(a)};
        \node at (5.7,0) {\includegraphics[width=0.25\textwidth]{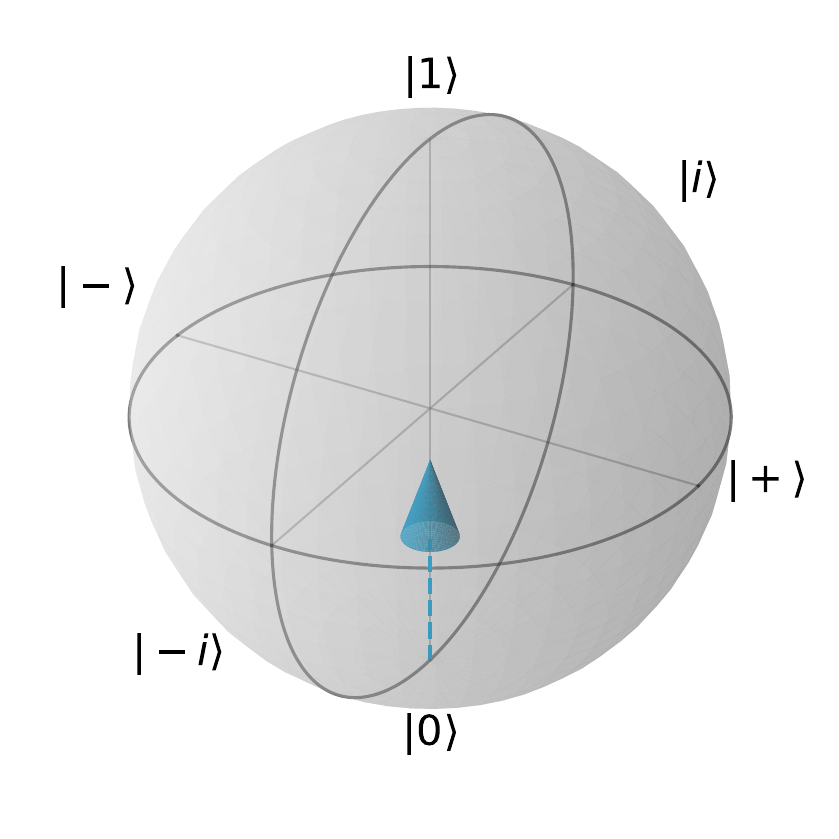}};
        \node at (4.2,1.8) {(b)};
        \node at (3.75,-3.8) {\includegraphics[width=0.25\textwidth]{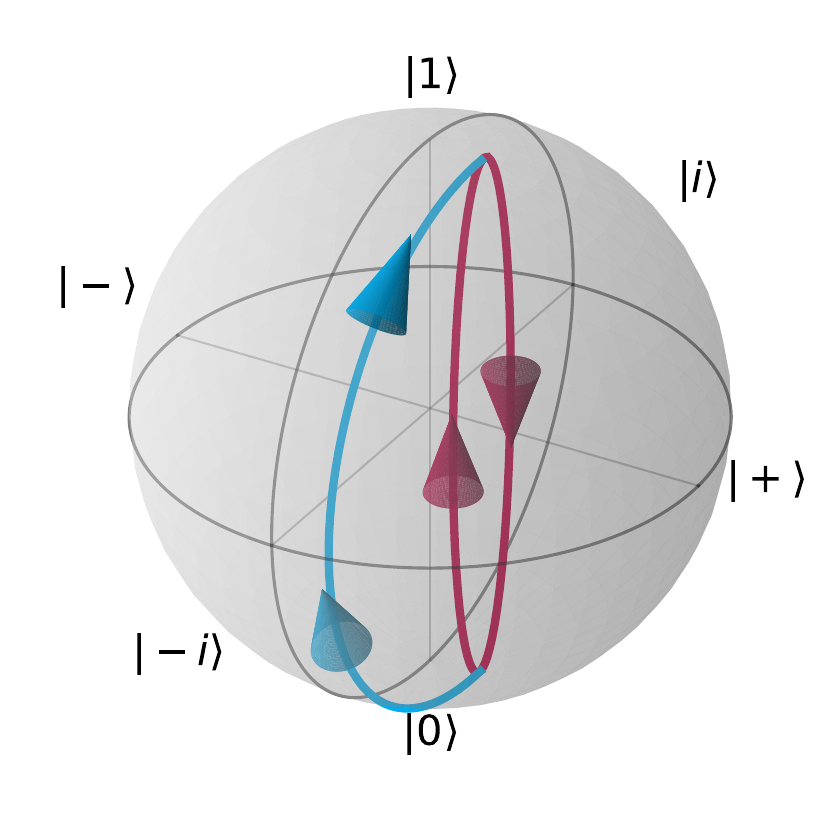}};
        \node at (2.25,-2) {(c)};
        \end{tikzpicture}
        \caption{\label{fig:battery} 
        Bloch-sphere representation of the first qubit of the battery when evolving freely (red) and when constrained to separable states (blue). 
        (a) Free evolution of the initial state $\ket{0}$ under $H_{\mathrm{D}}$. 
        Due to entanglement being generated during the evolution, the reduced state moves through the interior of the Bloch sphere.
        (b) Constrained evolution of the initial state $\ket{0}$. 
        Due to $H_{\mathrm{D}}$ driving orthogonal to the set of separable states (surface of the Bloch sphere), the system remains in the state $\ket{0}$.
        (c) Free and constrained evolution of the initial state $\ket{\psi(0)} \propto \ket{0}+\varepsilon\ket{1}$ for $\varepsilon=0.1$.
        The constrained state changes slower than the freely evolving one, and remains on the surface of the Bloch sphere as its global state is separable.
        }
    \end{figure}
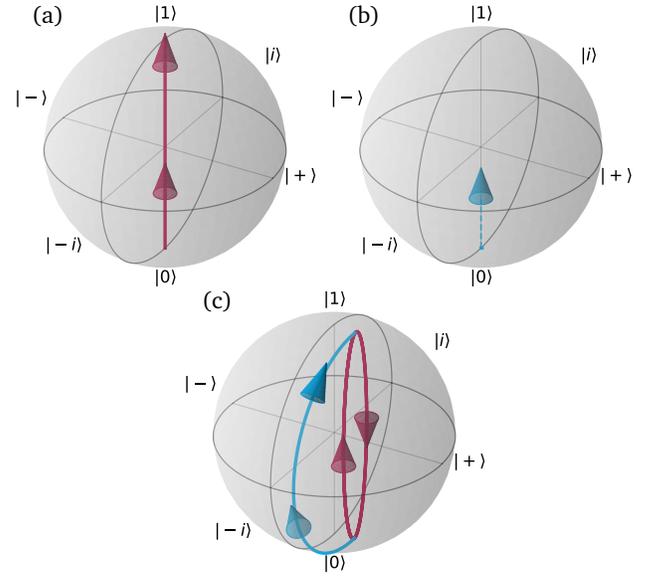

    \subsection{Charging a quantum battery via separable states}
    \label{sec:battery}

    Consider a quantum battery consisting of two qubits. 
    The system is initially in the ground state $\ket{00}$, and its internal Hamiltonian is $H_0 = \omega(\sigma^{z}\otimes \mathbb{1} + \mathbb{1} \otimes\sigma^{z})$, where $\sigma^{j}$ denotes the $j$th Pauli matrix.
    To bring the system into the state $\ket{11}$, that is, to charge the battery, the driving Hamiltonian $H_{\mathrm{D}} = \kappa \sigma^{x}\otimes\sigma^{x}$ is applied \cite{BVM15,AF13}. 
    This performs work on the system because its energy changes with respect to $H_0$.
    This can be seen more formally by moving into an interaction picture, $H_{\mathrm{D}}\mapsto e^{-i H_0 t} H_{\mathrm{D}} e^{i H_0 t}$, and using Eq.~\eqref{eq:heat-work}.
    
    Solving Schrödinger's equation for this system yields
    \begin{equation}
        \label{eq:state_battery}
        \ket{\psi(t)}=\cos\left(\kappa t\right)\ket{\psi_0}-i\sin\left(\kappa t\right)\sigma^x\otimes \sigma^x\ket{\psi_0}.
    \end{equation}
    The evolution is shown in Fig.~\ref{fig:battery} (a) in terms of the Bloch vector of the first qubit, with the joint system being initially in the state $\ket{\psi_0}=\ket{00}$.
    Due to the buildup of entanglement in the two-qubit state~~\eqref{eq:state_battery}, the first qubit is in a mixed state, thus passing through the interior of the Bloch sphere.

    To study the dynamics constrained to separable states, we seek a solution to the separability Schrödinger equations~\eqref{eq:SSE}.
    The reduced operators are $(H)_A=\kappa_B \sigma^x$ and $(H)_B=\kappa_A \sigma^x$, with couplings 
    \begin{equation}
        \label{eq:constrained-coupling}
        \begin{split}
            \kappa_k & = \kappa\braket{\psi_k|\sigma^x|\psi_k} = \kappa\left(\psi_k^0\right)^*\psi_k^1 +  \kappa\left(\psi_k^1\right)^*\psi_k^0
        \end{split}
    \end{equation}
    being proportional to the $x$ component of the local state $\ket{\psi_k(t)}=\psi_k^0(t)\ket{0} + \psi_k^1(t)\ket{1}$ of qubit $k=A,B$.
    The solution to Eq.~\eqref{eq:SSE} is obtained by direct integration, 
    {\begin{equation*}
        \begin{split}
            \ket{\psi_A(t)} & = \cos(K_B)\ket{\psi_A(0)} - i\sin(K_B)\sigma^x\ket{\psi_A(0)},\\
            \ket{\psi_B(t)} & = \cos(K_A)\ket{\psi_B(0)} - i\sin(K_A)\sigma^x\ket{\psi_B(0)},\\
        \end{split}
    \end{equation*}
    where $K_k(t)=\int^t\kappa_k(t^\prime)dt^\prime$. Despite the time dependence of $\ket{\psi_k(t)}$ in Eq.~\eqref{eq:constrained-coupling}, here we find $\braket{\psi_k (t)|\sigma^x|\psi_k (t)}=\braket{\psi_k(0)|\sigma^x|\psi_k(0)}$, so that $K_k=\kappa_k t$.}
    As intended, the constrained dynamics ensure that the system remains in a product state at all times $t$. 
    Nevertheless, the qubits still interact, owing to the fact that the coupling strength $\kappa_B$ experienced by qubit $A$ depends on the state of qubit $B$; see Eq.~\eqref{eq:constrained-coupling}.
    Note that for the initial state $\ket{\psi_k(0)}=\ket{0}$, for $k=A,B$, the constrained state does not evolve at all [Fig.~\ref{fig:battery} (b)] because $H_{\mathrm{D}}$ drives the state orthogonal to the set of separable states.
    This shows that the entire charging of the battery is caused by entanglement, not just a part of it. 
    For a slightly different initial state, e.g., $\ket{\psi_k(0)} \propto \ket{0}+\varepsilon\ket{1}$, the driving $H_{\mathrm{D}}$ is no longer orthogonal to the set of separable states.
    This allows for a partial charge of the battery, though at a slower rate $\kappa_k t\leq \kappa t$; see Fig.~\ref{fig:battery} (c). 
    Such slow-downs are a generic feature of constrained dynamics \cite{YS24}.

    \subsection{Markovian dynamics constrained to separable states}

    To study the impact of entanglement on a system's heat exchange, the constrained dynamics must be extended to open quantum systems.
    For a Markovian evolution, the freely evolving state $\rho(t)$ of a composite system $AB$ is governed by the master equation~\eqref{eq:LMEq}.
    For a small time step $\tau$, the solution of Eq.~\eqref{eq:LMEq} is given by the Kraus map \cite{MCD93}
    \begin{equation}
        \label{eq:FormalSolution}
          \rho(t+\tau)=\sum_{m} K_m(\tau)\rho(t)K_m^{\dag}(\tau).
	\end{equation}
    Here, $K_0=\mathbb{1}^{\otimes 2} - i\tau H-\frac{\tau}{2}\sum_k L^{\dag}_kL_k$
	describes a deterministic evolution that is occasionally interrupted by the quantum jumps $K_{m}=\sqrt{\tau}L_{m}$.
    Given the system to be initially in a product state, $\ket{\psi_A\psi_B}$, the state after a time $\tau$ corresponds to an ensemble of possibly entangled states $K_m\ket{\psi_A\psi_B}$.

    To constrain the open-system dynamics to separable states, we calculate the reduced Kraus operators
    \begin{equation}
        \begin{split}
            (K_0)_A & = \mathbb{1}{-}i\tau(H)_A{-}\tfrac{\tau}{2}\sum_k\big(L_k^{\dag}L_k\big)_A,\\
            (K_0)_B & = \mathbb{1}{-}i\tau(H)_B{-}\tfrac{\tau}{2}\sum_k\big(L_k^{\dag}L_k\big)_B,
        \end{split}
    \end{equation}
    and
    \begin{equation}
        (K_m)_A=\sqrt{\tau}(L_m)_A,\quad (K_m)_B{=}\sqrt{\tau}(L_m)_B,
    \end{equation}
    for subsystem $A$ and $B$, respectively.
    This approach removes any entanglement from the evolution, while preserving all classical correlations.
    The state after time $\tau$ reads \cite{PA24,AP24}
	\begin{equation}
        \label{eq:SepMap}
        \rho_\ms(t+\tau)=\sum_{m} \frac{\left(K_m\right)_A\otimes \left(K_m\right)_B}{\braket{K_{m}}}\rho_\ms(t)\frac{\left(K_m\right)_A^\dag\otimes \left(K_m\right)_B^\dag}{\braket{K_{m}}^*}.
	\end{equation}
    Comparing the evolved state in Eq.~\eqref{eq:SepMap} with $\rho_\ms(t+\tau)\approx \rho_\ms(t)+\tau \tfrac{d\rho_\ms}{dt}$, while keeping only contributions up to first order in $\tau$, yields
    \begin{equation}
        \label{eq:SepLin}
        \begin{split}
        \frac{d \rho_\ms}{d t} & =\mathcal{L}_\ms(\rho_\ms)\\
        & = i\left[\rho_\ms,\left(H\right)_A\otimes \mathbb{1}+\mathbb{1}\otimes \left(H\right)_B\right]+\sum_{k}\mathcal{D}_{\ms,k}(\rho_{\ms}).\\
        \end{split}
    \end{equation}
    Here, dissipation through separable states is due to
    \begin{equation*}
        \begin{split}
            \mathcal{D}_{\ms,k}(\rho_{\ms}) & = \frac{\left(L_k\right)_A\otimes \left(L_k\right)_B}{\braket{L_{k}}}\rho_\ms \frac{\left(L_k\right)_A^\dag\otimes \left(L_k\right)_B^\dag}{\braket{L_{k}}^*} + \braket{L_k^{\dag} L_{k}}\rho_\ms\\
            & \quad -\frac{1}{2}\big\{\big(L_k^{\dag} L_{k}\big)_A\otimes \mathbb{1}+\mathbb{1}\otimes\big(L_k^{\dag} L_{k}\big)_B,\rho_\ms\big\}.\\
        \end{split}
    \end{equation*}
    We refer to the nonlinear equation of motion~\eqref{eq:SepLin} as the separability Lindblad equation \cite{PA24}.
    It describes the evolution of a composite open quantum system restricted to a separable state. 
    The solution of Eq.~\eqref{eq:SepLin} can be compared with those of the conventional Lindblad equation to study the effect of entanglement on the thermodynamic properties of a system.
    If the freely evolving state $\rho$ does not become entangled during the process, then it coincides with constrained one at all times, i.e., $\rho_\ms(t)=\rho(t)$.
    Unlike for the conventional master equation~\eqref{eq:LMEq}, the solution to Eq.~\eqref{eq:SepLin} is generally not trace-preserving but can easily be made so by replacing the generator $\mathcal{L}_\ms(\rho_\ms)$ with $\mathcal{L}_\ms(\rho_\ms) -\Tr[\mathcal{L}_\ms(\rho_\ms)]\rho_\ms$, with the trace term accounting for normalization without changing correlations.
    
    Note that, Eq.~\eqref{eq:SepLin} is only defined for a pure product state, $\rho_\ms=\ket{\psi_A\psi_B}\bra{\psi_A\psi_B}$, since after an infinitesimal time step $\tau$, dissipation leads $\rho(t)$ to evolve into a mixture
    \begin{equation}
        \rho_\ms(t+\tau) = \rho_\ms + \tau \mathcal{L}_\ms(\rho_\ms) = \sum_m \rho_\ms^m(t+\tau).
    \end{equation}
    Each pure-state, $\rho_\ms^m(t+\tau) = \ket{\psi_A^m\psi_B^m}\bra{\psi_A^m\psi_B^m}$, with
    \begin{equation}
        \ket{\psi_A^m\psi_B^m} = \frac{\left(K_m\right)_A\otimes \left(K_m\right)_B}{\braket{K_{m}}}\ket{\psi_A\psi_B},
    \end{equation}
    has to be propagated separately by Eq.~\eqref{eq:SepLin} to obtain the state at a later time, viz.
    \begin{equation}
        \rho_\ms^m(t+2\tau)=\rho_\ms^m(t+\tau) + \tau\mathcal{L}_\ms[\rho_\ms^m(t+\tau)].
    \end{equation}
    Averaging then yields the state $\rho_s(t+2\tau)=\sum_m \rho^m_s(t+2\tau)$, and the iteration continues for later times.
    Therefore, Eq.~\eqref{eq:SepLin} is not a differential equation but an iteration prescription for quantum trajectories \cite{AP24}.

    In many practical situations, the number of possible trajectories becomes quickly unmanageable (even numerically).
    Therefore, we may randomly choose the state $\rho^m_\ms$ at a time step according to the probability $p_m=\Tr\rho^m_\ms$ and then average over many random realizations of this process.
    This is achieved by a Monte Carlo wave-function approach \cite{DZR92,MCD93,PK98} using the reduced Kraus operators in Eq.~\eqref{eq:SepMap}.
    While this approach is often viewed as an alternative to the solution of the conventional master equation~\eqref{eq:LMEq}, in our case it is an indispensable tool to carry out the iteration governed by Eq.~\eqref{eq:SepLin}.
    In Appendix~\ref{app:numerics}, we provide the algorithm for determining the constrained dynamics numerically.
    There, we also extend the above analysis to multipartite systems, as needed for the analysis of the three-qubit quantum refrigerator in Sec.~\ref{sec:fridge}.

    \section{Separable Quantum Thermodynamics}
    \label{sec:sep-thermo}

    For a system whose non-equilibrium dynamics are constrained to separable states, the first law~\eqref{eq:1st-law} reads
    \begin{equation}
        \label{eq:1st-law-sep}
        \begin{split}
            \dot{U}_\ms & = \dot{Q}_\ms + \dot{W}_\ms\\
            & = \Tr(\dot{\rho}_{\ms} H) + \mathrm{Tr}(\rho_{\ms} \dot{H}),\\
        \end{split}
    \end{equation}
    where heat and work are evaluated on the constrained state $\rho_{\ms}$. 
    For a Markovian environment, $\rho_{\ms}(t)$ is governed by the separability Lindblad equation~\eqref{eq:SepLin}. 
    Inserting Eq.~\eqref{eq:SepLin} for $\dot{\rho}_\ms$ in Eq.~\eqref{eq:1st-law-sep}, the constrained heat becomes
    \begin{equation}
        \label{eq:sep-heat}
            \dot{Q}_\ms = \Tr(\dot{\rho}_{\ms} H) = \sum_a \Tr[\mathcal{D}_{\ms,a}(\rho_{\ms}) H].
    \end{equation}
    Despite the fact that the constrained Hamiltonian $H_\ms$ in Eq.~\eqref{eq:constrained-H} does not need to commute with the free Hamiltonian, $[H_\ms,H]\neq 0$, there is no unitary contribution to the heat flow in Eq.~\eqref{eq:sep-heat}, viz. 
    \begin{equation}
        \begin{split}
            \Tr([\rho_\ms,H_\ms]H) & = \bra{\psi_A\psi_B}(H_\ms H - H H_\ms )\ket{\psi_A\psi_B}\\
            & = 0.\\
        \end{split}
    \end{equation}
    The first equality is obtained from $\rho_\ms = \ket{\psi_A\psi_B}\bra{\psi_A\psi_B}$ and the second equality follows from
    \begin{equation*}
        \begin{split}
            \bra{\psi_A\psi_B}H_\ms H\ket{\psi_A\psi_B} & = \bra{\psi_A\psi_B}H\ket{\psi_A}\bra{\psi_A}H\ket{\psi_A\psi_B}\\
            &\quad + \bra{\psi_A\psi_B}H\ket{\psi_B}\bra{\psi_B}H\ket{\psi_A\psi_B}\\
            & = \bra{\psi_A\psi_B}H H_\ms\ket{\psi_A\psi_B},\\
        \end{split}
    \end{equation*}
    where we used Eq.~\eqref{eq:red-ops}.

    \subsection{Constrained entropy production}

    To characterize the irreversible nature of the constrained dynamics, we consider a system that is coupled to a thermal bath such that the resulting evolution is Gibbs preserving \cite{FOR15}, i.e., $\mathcal{L}_t[\gamma(t)]=0$, where $\gamma(t)\propto \exp[-\beta H(t)]$ is a thermal state and $H(t)$ is the system's Hamiltonian.

    The constrained evolution from $\rho_\ms(t)$ to $\rho_\ms(t+\tau)$ follows the dynamical map~\eqref{eq:SepMap}.
    {If there exists an instantaneous steady state $\gamma_\ms(t)$, i.e., $\mathcal{L}_{\ms}(\gamma_\ms)=0$, where $\mathcal{L}_{\ms}$ is the generator~\eqref{eq:SepLin} obtained from the state $\rho_{\ms}(t)$ at time $t$,
    then} monotonicity of the relative entropy~\eqref{eq:ent-ineq} implies
    \begin{equation}
        \label{eq:H-theorem}
        S[\rho_\ms(t+\tau)||\gamma_{\ms}] \leq S[\rho_\ms(t)||\gamma_{\ms}].
    \end{equation}
    Considering $\tau$ to be small and writing $\rho_\ms(t+\tau)=\rho_\ms(t)+\tau \hat{\mathcal{L}}_\ms[\rho_\ms(t)]$ using Eq.~\eqref{eq:SepLin} yields 
    \begin{equation}
        S[\rho_\ms+\tau \hat{\mathcal{L}}_\ms(\rho_\ms)||\gamma_{\ms}] - S[\rho_\ms||\gamma_{\ms}] \leq 0.
    \end{equation}
    Here, $\hat{\mathcal{L}}_\ms(\rho_\ms) = \mathcal{L}_\ms(\rho_\ms) -\Tr[\mathcal{L}_\ms(\rho_\ms)]\rho_\ms$ yields a trace-preserving form of Eq.~\eqref{eq:SepLin}.
    Using Eq.~\eqref{eq:rel-ent}, we find a nonnegative entropy production rate,
    \begin{equation}
        \label{eq:ent-prod-sep}
        \begin{split}
            \sigma_\ms(t) & = - \frac{d}{dt}S[\rho_\ms(t)||\gamma_{\ms}]\geq 0.\\
        \end{split}
    \end{equation}
    Equation~\eqref{eq:ent-prod-sep} represents a kind of $H$-theorem \cite{S78} and is the closest we come to a second-law(-like) expression for the constrained non-equilibrium dynamics.
    {Note that the existence of the constrained steady state $\gamma_\ms$ is a sufficient condition for Eq.~\eqref{eq:ent-prod-sep} to hold.
    In general, nonlinear dynamical maps do not obey the monotonicity in Eq.~\eqref{eq:ent-ineq}.}

    {The above discussion} has an interesting consequence for the second law of thermodynamics: 
    if the Hamiltonian $H$ is able to create entanglement between different constituents of the system, then the thermal state $\gamma\propto \exp(-\beta H)$ can itself be entangled.
    Usually this is the case for small temperatures $\beta^{-1}$, where $\gamma$ can be approximated by a statistical mixture of only a few (entangled) eigenstates of $H$. 
    Since the constrained state $\rho_{\ms}(t)$ remains separable, it cannot converge to this equilibrium state, i.e., $\rho_{\ms}(t\to\infty)=\gamma_\ms\neq \gamma$.
    Then, it is not possible to write Eq.~\eqref{eq:ent-prod-sep} as a balance equation~\eqref{eq:2nd-law}.
    Informally speaking, a thermal bath does not necessarily thermalize the system when constrained to separable states.

    Note that even in the case of the Hamiltonian being local, $H=H_\ms$, and thus the thermal state $\gamma$ being separable, the steady state $\gamma_\ms$ of the constrained evolution can be different, $\gamma_\ms\neq \gamma$. 
    This is possible because entanglement can be created by the jump operators $L_k$ corresponding, for instance, to correlated decay channels \cite{KRS11,TCM11,CEZ22}. 
    The action of some $L_k$ might move the state orthogonal to the set of separable states and this effect is removed for the constrained dynamics; recall Fig.~\ref{fig:battery} (b).

    \section{Heating and cooling from entanglement}
    \label{sec:apply}

    How much heat would a quantum system accumulate without entanglement?
    This counterfactual cannot be answered by quantifying the amount of entanglement during the evolution and comparing it to the heat.
    An active intervention is necessary that constrains the specific system under investigation to evolve through separable states.
    Comparing the constrained trajectories with the freely evolving state, we are able to identify how entanglement caused the heat of a quantum system to change. 
    Here, we illustrate this for a three-qubit model of a quantum refrigerator and the dephasing of a two-qubit system with respect to an entangled basis.
    Furthermore, we use the constrained dynamics to remove system-bath entanglement from the interaction of a qubit with its environment and calculate the resulting heat exchange.

    \subsection{Entanglement in quantum refrigerators}
    \label{sec:fridge}
    
    The simplest model of a quantum refrigerator consists of three qubits that we label as $\mathrm{w}$ (work), $\mathrm{h}$ (hot), and $\mathrm{c}$ (cold) \cite{LPS10, CPA13, SBLP11}.
    Their respective energies are $\omega_\mathrm{w}$, $\omega_\mathrm{h}$ and $\omega_\mathrm{c}$. 
    The energies are such that $\omega_\mathrm{h}=\omega_\mathrm{c}+\omega_\mathrm{w}$, so excitations of the hot qubit can be interchanged by excitations of the cold and work qubits without the need for external energy. 
    The free Hamiltonian of the system is
    \begin{equation}
        H_0 = \omega_\mathrm{w} \ketbra{1}{1}_\mathrm{w} +  \omega_\mathrm{h} \ketbra{1}{1}_\mathrm{h} +  \omega_\mathrm{c} \ketbra{1}{1}_\mathrm{c},
    \end{equation}
    where $\ketbra{1}{1}_\mathrm{w}=\ketbra{1}{1}\otimes\mathbb{1}^{\otimes 2}$ for the qubit $\mathrm{w}$ and similarly for $\mathrm{h}$ and $\mathrm{c}$.
    Coherent transfer of energy between the qubits is caused by \cite{CPA13}
    \begin{equation}
        H_{\mathrm{I}} = g\big(\ketbra{101}{010}+\ketbra{010}{101}\big).
    \end{equation}
    Additionally, each qubit is coupled to a thermal bath at temperature $T_\mathrm{w}$, $T_\mathrm{h}$, and $T_\mathrm{c}$, leading to thermalization of each individual qubit \cite{SBLP11}, viz.
    \begin{equation}
    \label{eq:local-dissipator}
        \mathcal{D}_k(\rho)=p_k[\tau_k\Tr_k(\rho)-\rho],
    \end{equation}
    where $\tau_k=e^{-E_k/T_k\ketbra{1}{1}_k}/(1+e^{-E_k/T_k})$ is the thermal state of the $k$th qubit, for $k=\mathrm{w},\mathrm{h},\mathrm{c}$.
    
\begin{figure}[t]
        \centering
        \includegraphics[width=0.48\textwidth]{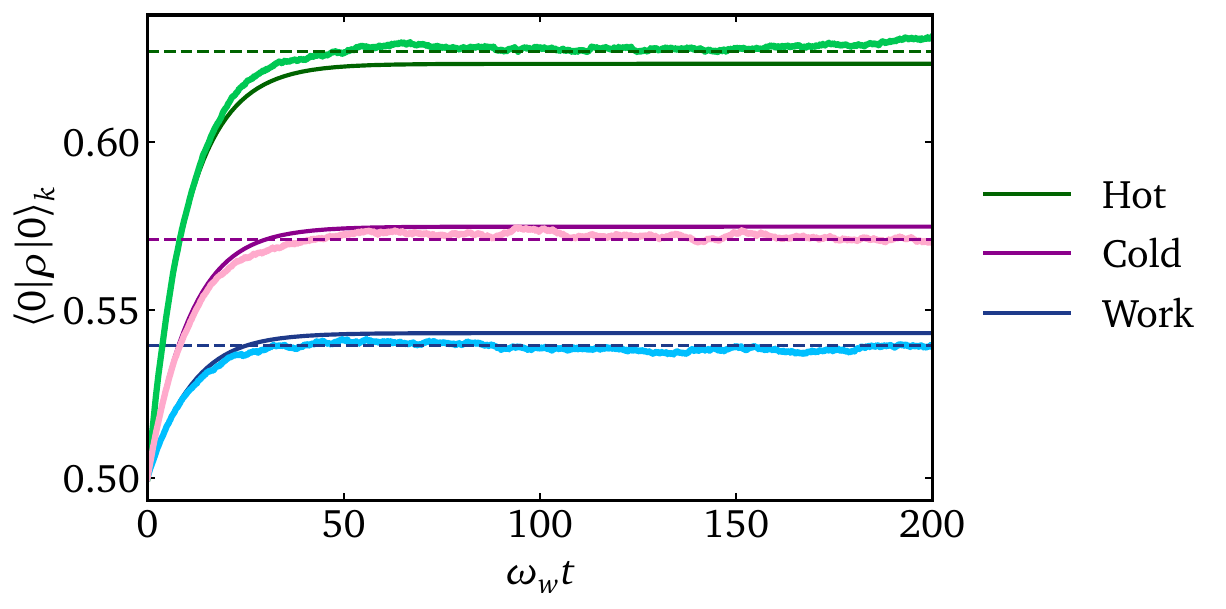}
        \caption{\label{fig:population-localized} Ground-state population of each qubit of the quantum refrigerator when coupled to individual thermal baths and for a system initialized in the state $\ket{+++}$.
        The constrained populations (light) approach the dashed lines, indicating the same populations as if $g=0$. 
        In contrast, the freely evolving populations (dark) approach values above or below these values, which is indicative of cooling or heating beyond the temperature of the bath. 
        Parameters are $T_{\mathrm{w}}=6.33$, $T_{\mathrm{h}}=3.25$, $T_{\mathrm{c}}=2.4$, $\omega_{\mathrm{w}}=1$, $\omega_{\mathrm{c}}=0.687$, $g=0.1$, and $p_i=0.1$. 
        The constrained populations are obtained by the Monte Carlo wave function method averaging over $4\times 10^6$ trajectories \cite{Data_set}.
        }
    \end{figure}

    For any initial state, the cold qubit $\mathrm{c}$ can be cooled below the temperature $T_{\mathrm{c}}$ of its environment \cite{LPS10}; see Fig.~\ref{fig:population-localized}. 
    As shown in Fig.~\ref{fig:population-localized}, the ground-state population of the cold and hot qubits is reduced (darker lines) below what would have been obtained without interaction (dashed lines), i.e., in the case $g=0$.
    It has therefore been claimed that this cooling advantage is to be attributed to entanglement generated by $H_{\mathrm{I}}$ \cite{BHL14}.
    However, this argument is not conclusive as the cooling advantage could be similarly attributed to the coherences, necessarily also generated during the coherent interaction via $H_{\mathrm{I}}$.
    Remarkably, the separable dynamics reveal that indeed all cooling advantage is due to entanglement as can be seen from the constrained ground-state populations (lighter lines in Fig.~\ref{fig:population-localized}).
    These populations approach the one of the non-interaction model, $g=0$, showing that having no entanglement leads to the same populations as having no interaction at all.

    \subsection{Delocalized model of thermalization}

    \begin{figure}[t]
        \centering
        \includegraphics[width=0.39\textwidth]{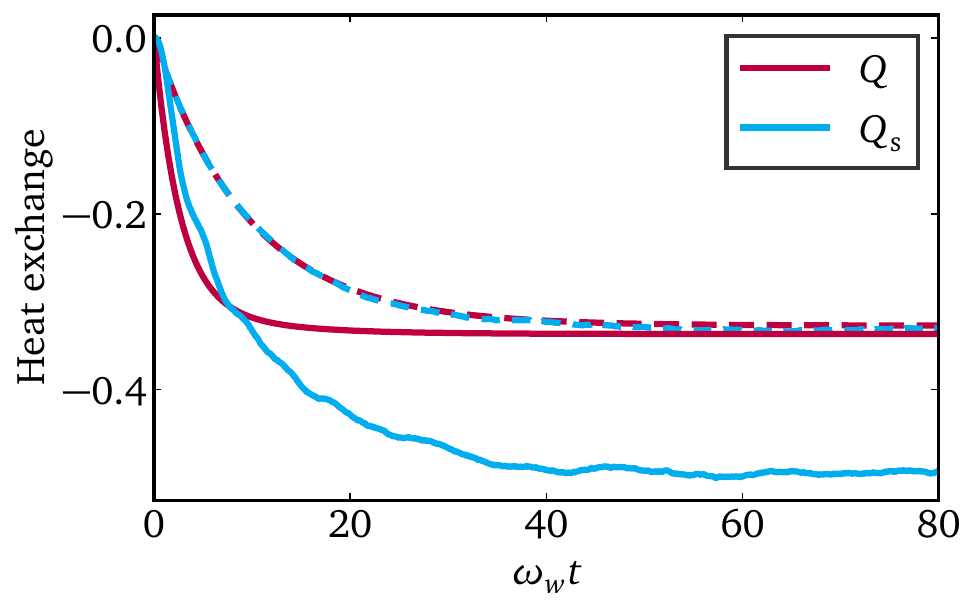}
        \caption{\label{fig:heat-fridges} Heat exchanged with the environment for a quantum refrigerator with (blue, $Q_s$) and without (red, $Q$) constraints, for the localized (dashed) and the delocalized (full) models. 
        With the initial state $\ket{+++}$, one observes larger disagreement between the heat flows for the delocalized model than for the localized one. 
        The parameters used are $T_\mathrm{w}=6.33$, $T_\mathrm{h}=3.25$, $T_\mathrm{c}=2.4$, $\omega_\mathrm{w}=1$, $\omega_\mathrm{c}=0.687$, $g=0.1$, and (a) $p_k=0.1$ and (b) $\gamma=0.01$. 
        The plots are the result of averaging $4\times10^6$ (localized) and $5\times10^5$ (delocalized) trajectories \cite{Data_set}.}
    \end{figure}

    In the above model, each qubit was coupled to a separate thermal bath leading to localized thermalization~\eqref{eq:local-dissipator}.
    Starting from the microscopic model of the qubits being coupled to bosonic baths, it was shown that localized thermalization demands a negligible interaction, $H_{\mathrm{I}}\approx 0$ \cite{CPA13}. 
    A consistent treatment of the full system-bath interaction leads to a delocalized model of thermalization for the refrigerator; that is, the dissipator in Eq.~\eqref{eq:local-dissipator} is replaced by 
    \begin{equation}
        \label{eq:deloc-dissipator}
        \mathcal{D}_k(\rho)=\sum_{\omega}\Gamma_{k,\omega}\Big(A_{k,\omega}\rho A_{k,\omega}^\dagger-\frac{1}{2}\{A_{k,\omega}^\dagger A_{k,\omega},\rho\}\Big),
    \end{equation}
    where $\omega = \pm\omega_k,\pm\omega_k\pm g$. 
    The damping rates are $\Gamma_{k,\omega}=\gamma\omega^3e^{\beta_k\omega/2}\sinh(\beta_k\omega/2)$ and the operators $A_{k,\omega}$ are listed explicitly in Appendix~\ref{app:delocalized} as well as Ref.~\cite{CPA13}.
    Unlike the case of local dissipation~\eqref{eq:local-dissipator}, these operators generate entanglement for $g>0$, e.g., $A_{\mathrm{w},\omega}\neq L_{\omega}\otimes \mathbb{1}^{\otimes 2}$, where $L_{\omega}$ is an operator acting only on the single qubit $\mathrm{w}$.

    Figure~\ref{fig:heat-fridges} shows the total heat exchange $Q(t)$ of the three-qubit system for both the localized and delocalized model (red lines). 
    The latter is calculated by integrating the heat flow $\dot{Q} = \sum_k\Tr[\mathcal{D}_k(\rho) (H_0+H_{\mathrm{I}})]$ up to time $t$ for the dissipator \eqref{eq:local-dissipator} and \eqref{eq:deloc-dissipator}, respectively. 
    We observe that the delocalized model achieves an enhanced cooling compared to the localized model.
    Additionally, Fig.~\ref{fig:heat-fridges} contains the accumulated heat $Q_\ms$ when the evolution is constrained to separable states (blue lines).  
    The latter is calculated using $\dot{Q}_\ms = \sum_k\Tr[\mathcal{D}_{k,\ms}(\rho_\ms) (H_0+H_{\mathrm{I}})]$ using the constrained operators. 
    While the constrained heat of the localized model differs only slightly from the unconstrained one, the delocalized model involves jump operators $A_{k,\omega}$ that can create entanglement, thus leading to more pronounced differences between the constrained and unconstrained evolution.
    
    \subsection{Quantum decoherence and pure dephasing}
    \label{sec:decoherence}

    There are certain processes which preserve a system's internal energy $E=\Tr(\rho H)$, i.e., $\dot{E}=0$.
    If the Hamiltonian $H$ of the system is time independent, $\dot{H}=0$, then it follows from Eq.~\eqref{eq:1st-law} that there is no heat exchange with the environment, $\dot{Q}=0$.
    A large class of Markovian processes that do not generate heat are those for which the Hamiltonian $H$ commutes with its jump operators, i.e., $[L_j,H]=0$, for all $j$.
    Note that $L_j H = H L_j$ implies $[L_j^\dag,H]=0$ as well.
    Using Eq.~\eqref{eq:LMEq} together with these commutation relations, one verifies that the heat flow,
    \begin{equation}
        \label{eq:no-heat-cond}
        \begin{split}
            \dot{Q} & = \Tr(\dot{\rho}H)\\
            & = \Tr(L_j \rho L_j^\dag H - \frac{1}{2}L_j^\dag L_j \rho H - \frac{1}{2}\rho L_j^\dag L_j H)\\
            & = 0,\\
        \end{split}
    \end{equation}
    does indeed vanish.
    Equation~\eqref{eq:no-heat-cond} shows that there is no population transfer between the system and the environment.
    The most prominent example of such a process is quantum decoherence \cite{Z91,S04}.
    Decoherence is a phenomenon by which the environment continuously monitors the system, similarly to a measurement apparatus.
    When the monitored observable is the system's Hamiltonian $H$, the jump operators $L_k = \ket{\phi_k}\bra{\phi_k}$ describe a projective measurement of the energy eigenstates $\ket{\phi_k}$.
    As time increases, the state $\rho(t)$ approaches a matrix that is diagonal with respect to the eigenbasis $\ket{\phi_k}$ while its off-diagonal elements are increasingly suppressed.
    Since $H\ket{\phi_k}\propto \ket{\phi_k}$, we have $[L_k,H]=0$, for all $k$.
    It follows from Eq.~\eqref{eq:no-heat-cond} that decoherence occurs without heat exchange, $\dot{Q}=0$.
    
    If an energy-preserving process~\eqref{eq:no-heat-cond} creates entanglement during the evolution of a bipartite system $AB$, then the constrained state $\rho_\ms$ will deviate from the freely evolving state $\rho$.
    In particular, the partially reduced operators $(L_k)_A$ and $(L_k)_B$ do not need to commute with $H$, even if $[L_k,H]=0$.
    In this case, it is possible to observe a non-vanishing heat $Q_\ms$ for the constrained evolution.
    In conclusion, quantum entanglement can suppress heat exchange during decoherence.

\begin{figure}[t]
        \centering
        \includegraphics[width=0.46\textwidth]{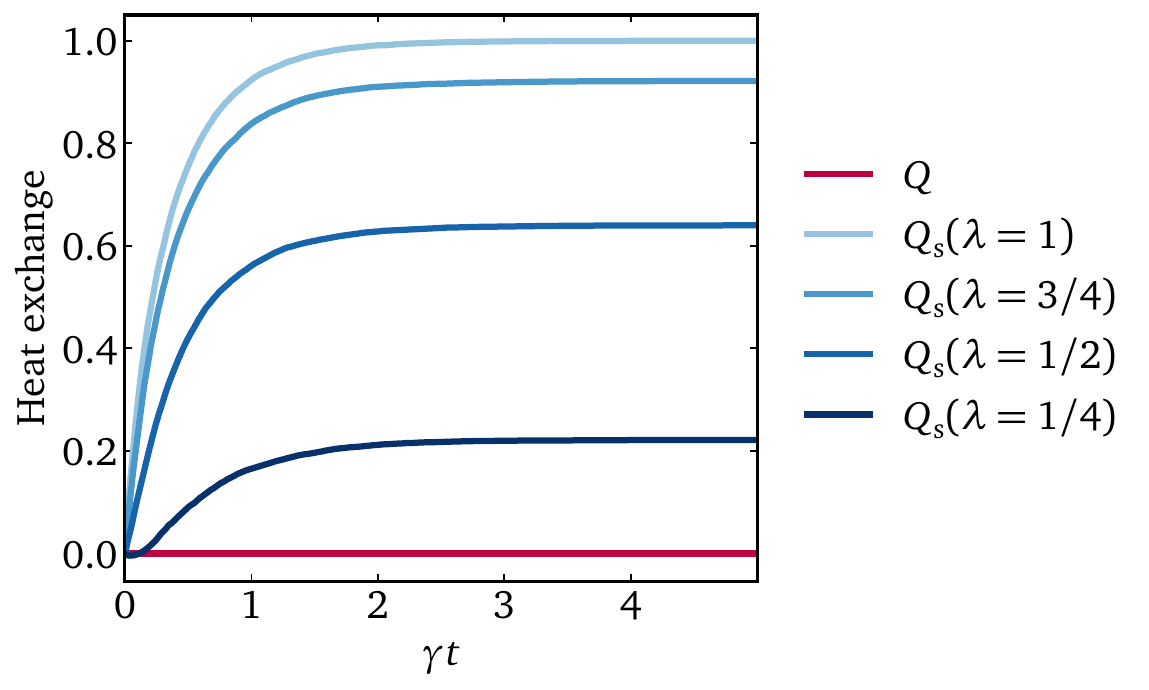}
        \caption{\label{fig:dephasing} 
        Heat exchange during a correlated dephasing process for the initial state $\ket{\psi_0}\propto (\ket{0}+\lambda\ket{1})^{\otimes 2}$, for $\lambda=1$, $3/4$, $1/2$, $1/4$, and $\omega=\gamma$. 
        The freely evolving system does not accumulate heat, $Q=0$. 
        In contrast, the evolution constrained to separable states experience a nonzero heat exchange, $Q_\ms>0$.
        Results were obtained via the Monte Carlo wave function method averaging $2\times10^4$ constrained trajectories for each $\lambda$
        \cite{Data_set}.}
    \end{figure}

    As an example, consider a two-qubit system that experiences correlated dephasing, $L = \sqrt{\gamma}(\sigma^z\otimes\mathbb{1} + \mathbb{1}\otimes\sigma^z)$, and is subject to a spin-spin interaction, $H = \omega \sigma^z\otimes \sigma^z$.
    Clearly, $[H,L]=0$, and thus the process obeys Eq.~\eqref{eq:no-heat-cond}.
    Hence, the freely evolving system does not accumulate heat, $Q=0$. 
    For the initial state $\ket{\psi_0}=\ket{+}^{\otimes 2}$, with $\ket{+}=(\ket{0}+\ket{1})/\sqrt{2}$, the Hamiltonian $H$ generates entanglement between the qubits.
    Constraining the non-equilibrium dynamics of the system to separable states removes the possibility of evolving through entangled states, and this constrained evolution may exhibit an exchange of energy with the environment.
    Figure~\ref{fig:dephasing} shows the accumulated heat $Q$ and $Q_\ms$ for the freely evolving and the constrained dynamics, respectively.
    During the constrained evolution $\rho_\ms(t)$ heat is accumulated, which is absent for the freely evolving state $\rho(t)$, allowing us to attribute the heat $Q_\ms$ to the absence of entanglement.
    
    \begin{figure*}[t]
        \centering
        \begin{tikzpicture}
        \node at (0,0) {\includegraphics[width=0.4\textwidth]{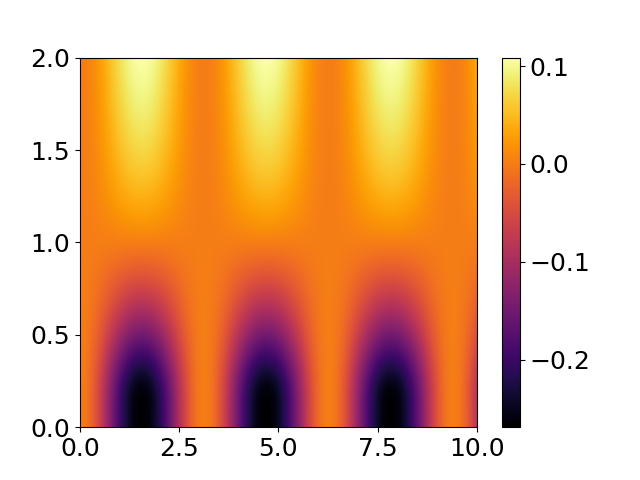}};
        \node at (8,0) {\includegraphics[width=0.4\textwidth]{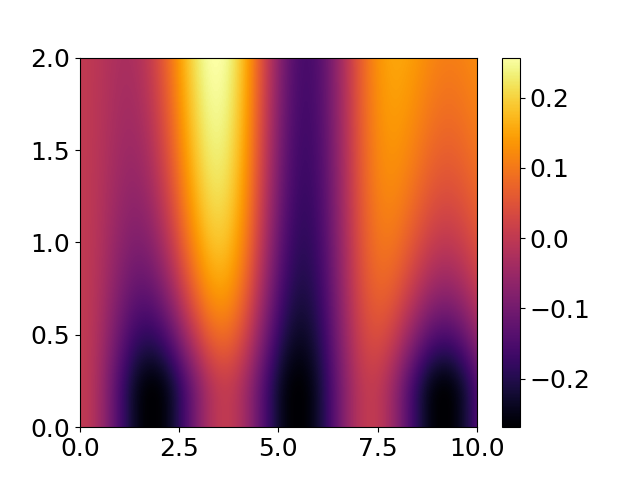}};
        \node at (-0.4,2.4) {Physical heat};
        \node at (2.15,2.4) {$Q$};
        \node at (-0.4,-2.8) {$t/\kappa$};
        \node at (7.6,2.4) {Constrained heat};
        \node at (10.15,2.4) {$Q_{\ms}$};
        \node at (7.6,-2.8) {$t/\kappa$};
        \node[rotate=90] at (4.4,0) {$T_B/T_A$};
        \node[rotate=90] at (-3.6,0) {$T_B/T_A$};
        \node at (-3.5,2.4) {(a)};
        \node at (4.5,2.4) {(b)};
        \node at (-3.5,-3.5) {(c)};
        \node at (4.5,-3.5) {(d)};
        \node at (-1,-6) {\includegraphics[width=0.35\textwidth]{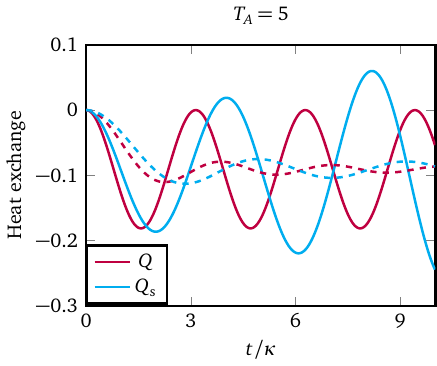}};
        \node at (7.1,-6) {\includegraphics[width=0.35\textwidth]{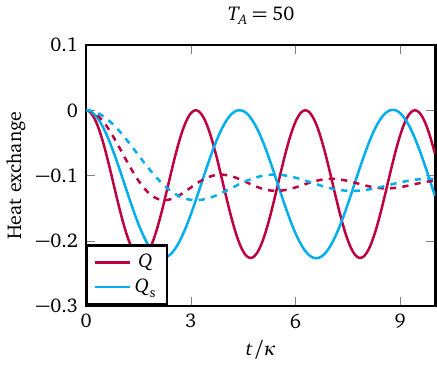}};
        \end{tikzpicture}
        \caption{\label{fig:system-bath-ent} (a) Heat $Q$ of qubit $A$ as a function of time $t$ and the quotient $T_B/T_A$, with $T_A=1$. 
        (b) Heat $Q_{\ms}$ for a time evolution constrained to separable states.
        (c),(d) Physical heat $Q$ and constrained heat $Q_{\ms}$ as a function of time $t$ (solid lines) as well as their time averages $\overline{Q}(t)$ and $\overline{Q}_\ms(t)$ (dashed lines), for $T_A\in\{5,50\}$ and $T_B=1$.
        In all plots, we set $\omega_A=\omega_B$.
        }
    \end{figure*}
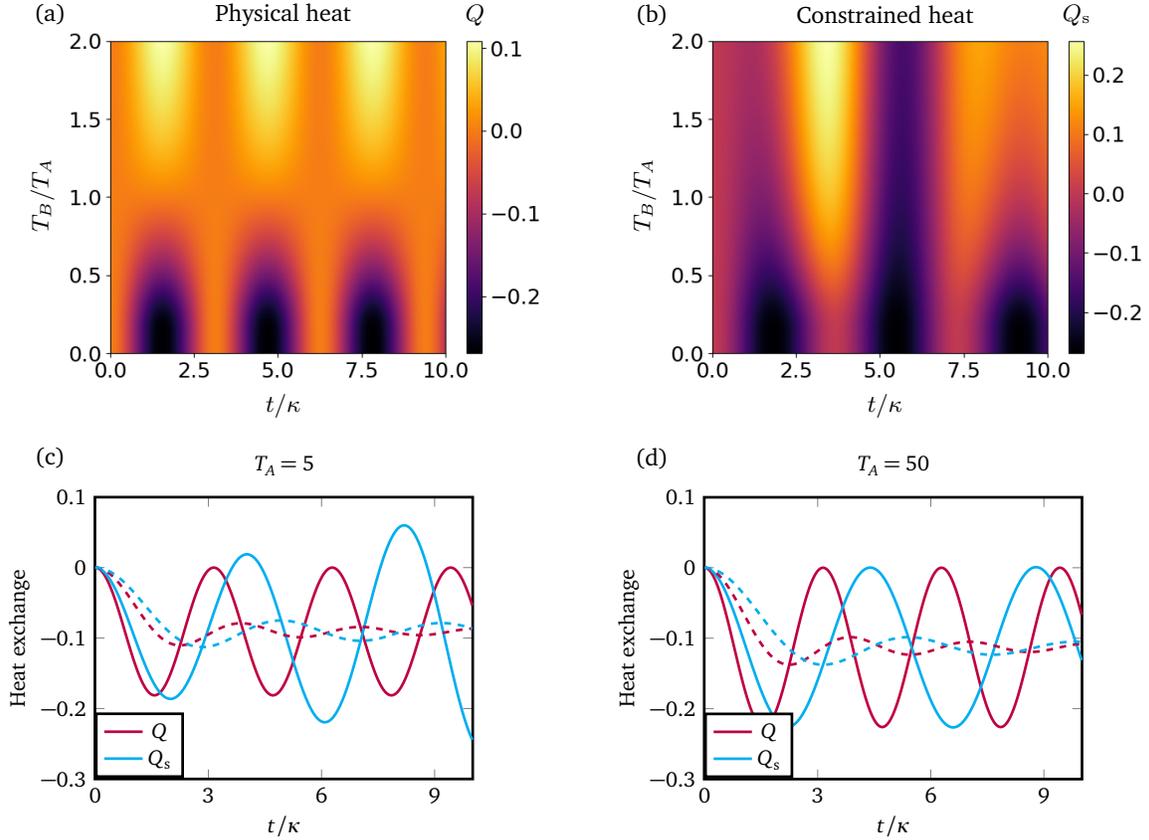

    \subsection{Heat exchange without system-bath entanglement}
    \label{sec:syst-bath-ent}

    So far the dynamics of a compound system were constrained such that entanglement between its constituents is removed.
    This pertains to possible advantages that quantum thermodynamic machines may offer over their classical counterparts. 
    Another type of entanglement which influences thermodynamic properties is found in system-environment correlations.
    These are unavoidable in any realistic physical setup and contribute to a system's change in internal energy.
    To remove system-environment entanglement, a constraint can be imposed on the unitary time evolution of the joint system-environment complex.
    For such a constraint, the generation of heat is governed by classical correlations with the environment.

    As an example, we study the heat exchange of a qubit $A$ due to its interaction with a two-level environment $B$.
    The system is prepared in the coherent superposition $\ket{\psi_A}\propto\ket{0}+ e^{-\beta_A \omega_A/2}\ket{1}$, while the environment is in a thermal state $\rho_B\propto\exp(-\beta_B H_B)$ of temperature $1/\beta_B$, with $H_{B}=\omega_B\ket{1}\bra{1}$.
    The qubit and environment interact according to the Hamiltonian $H=\kappa V$, where $V$ is the swap operator, $V\ket{\psi_A\psi_B}=\ket{\psi_B\psi_A}$.
    The heat is calculated from Eq.~\eqref{eq:heat} leading to
    \begin{equation}
        \label{eq:phys-heat-TLS}
        \begin{split}
            Q(t) & = \Tr\{H_A[\rho_A(t) - \ket{\psi_A}\bra{\psi_A}]\}\\
            & = \omega_A\sin^2(\kappa t)\left(\frac{e^{-\beta_B \omega_B}}{1 + e^{-\beta_B \omega_B}}-\frac{e^{-\beta_A\omega_A}}{1 + e^{-\beta_A\omega_A}}\right),\\
        \end{split}
    \end{equation}
    where $H_A=\omega_A\ket{1}\bra{1}$ is the free Hamiltonian of the system and $\rho_A(t)=\Tr_B\rho(t)$ is its reduced state, with $\rho_A(0)=\ket{\psi_A}\bra{\psi_A}$.
    See Appendix~\ref{app:exchange} for details.

    Next, consider an evolution constrained to separable states.
    The constrained dynamics~\eqref{eq:SSE} are determined by the reduced operators, 
    \begin{equation*}
        (V)_A=\ket{\psi_B(t)}\bra{\psi_B(t)},\quad (V)_B=\ket{\psi_A(t)}\bra{\psi_A(t)}.
    \end{equation*}
    Equation~\eqref{eq:SSE} can be solved analytically \cite{SW17},
    \begin{equation}
        \label{eq:sep-sol}
        \begin{split}
            \ket{\psi_A(t)} & = \cos(|q|\kappa t)\ket{\psi_A} - i\frac{q^*}{|q|}\sin(|q|\kappa t)\ket{\psi_B},\\
            \ket{\psi_B(t)} & = \cos(|q|\kappa t)\ket{\psi_B} - i\frac{q}{|q|}\sin(|q|\kappa t)\ket{\psi_A},\\
        \end{split}
    \end{equation}
    with $q=\braket{\psi_A|\psi_B}$ being the overlap between the initial states $\ket{\psi_A}$ and $\ket{\psi_B}$.
    Since the environment starts out in a mixed state, $\rho_B$, we have
    two initial states, $\ket{\psi_B}=\ket{0}$ and $\ket{\psi_B}=\ket{1}$, which have to be propagated individually using Eq.~\eqref{eq:sep-sol}.
    The constrained heat acquired during the evolution is given by Eq.~\eqref{eq:heat} and reads
    \begin{equation}
        \label{eq:sep-heat-TLS}
        \begin{split}
            Q_{\ms}(t) & = \Tr\{H_A(\rho_{\ms}(t)-\ket{\psi_A}\bra{\psi_A})\}\\ 
            & = \omega_Ae^{-\beta_A\omega_A}\frac{\cos^2(q_0\kappa t) + \cos^2(q_1\kappa t)e^{-\beta_B\omega_B}}{\big(1 + e^{-\beta_A\omega_A}\big)\big(1 + e^{-\beta_B\omega_B}\big)}\\
            &\quad +\frac{\omega_Ae^{-\beta_B\omega_B}}{1 + e^{-\beta_B\omega_B}}\sin^2(q_1\kappa t)-\frac{\omega_A e^{-\beta_A\omega_A}}{1 + e^{-\beta_A\omega_A}},\\
        \end{split}
    \end{equation}
    where we have $q_0=\braket{0|\psi_A}=1/\sqrt{Z_A}$ and $q_1=\braket{1|\psi_A}=e^{-\beta_A\omega_A/2}/\sqrt{Z_A}$, and $Z_A=1+e^{-\beta_A \omega_A}$.
    We refer to Appendix~\ref{app:exchange} for further details.
    
    Figure~\ref{fig:system-bath-ent} (a) shows the heat $Q$ accumulated during the process and Fig.~\ref{fig:system-bath-ent} (b) shows the constrained heat $Q_\ms$.
    While the heat $Q$ shows a clear separation at $T_A=T_B$ between the qubit being cooled or heated by the environment, the constrained heat $Q_{\ms}$ shows both regions of cooling and heating for both $T_A\geq T_B$ and $T_A\leq T_B$.
    The constrained heat $Q_\ms$ shows an additional heating and cooling of the system compared to $Q$, though at a smaller frequency, because the constrained evolution is slower in swapping the system's state $\ket{\psi_A}$ with the thermal state $\rho_B$; see also Ref.~\cite{YS24}. 
    This can be seen in Figs.~\ref{fig:system-bath-ent} (c) and (d), which show $Q(t)$ and $Q_{\ms}(t)$ (solid lines) for different initial coherences $T_A$ of the state $\ket{\psi_A}$.
    As $T_A$ gets larger, the initial state $\ket{\psi_A}$ approaches the superposition $\ket{+}=(\ket{0}+\ket{1})/\sqrt{2}$.
    Then, we also have that $q_0\approx q_1$, so that both (constrained) trajectories oscillate with the same frequency.
    In general, $q_1 \leq q_0$ and therefore the constrained state, unlike the freely evolving one, cannot return after a single oscillation into the thermal state $\rho_B$.
    Since a thermal state $\rho_B$ minimizes the (free) energy of the system \cite{JR10}, the constrained evolution shows an amplified exchange of energy with the environment.

    For the finite-dimensional environment $B$ no effective steady state exists as there is a (non-Markovian) revival of coherences in $\rho_A(t)$ caused by the interaction $\kappa V$.
    Nevertheless, we can factor in repeated oscillations in the heat $Q(t)$ by evaluating the time-averaged heat exchange $\overline{Q}(t) = \tfrac{1}{t}\int_{0}^t Q(t^\prime) dt^\prime$, and analogously $\overline{Q}_\ms(t)$ for the constrained heat.
    These time averages are shown in Figs.~\ref{fig:system-bath-ent} (c) and (d) (dashed lines).
    Although system-bath entanglement has a significant impact on the non-equilibrium heat exchange, the differences between $Q$ and $Q_\ms$ average out in the mean long-time equilibrium heat, i.e., $\overline{Q}(t\to\infty)=\overline{Q}_\ms(t\to \infty)$. 

    \section{Conclusion}
    \label{sec:fin}
    
    In this paper, we studied the non-equilibrium thermodynamics of composite quantum systems constrained to separable states.
    This allowed us to unambiguously identify the influence entanglement has on the non-equillibrium thermodynamics of a system, such as work and heat.
    Our theory applies to arbitrary multipartite systems and can be used to constrain system-system and system-bath entanglement between arbitrary partitions of its subsystems.
    
    We derived that the first and second laws of quantum thermodynamics hold true even for a constrained system. 
    The benefits of our framework was illustrated through several benchmark systems. 
    For a two-qubit battery, we found that without entanglement, the charging process is dramatically slowed down.
    Moreover, we studied the role of entanglement in different models of quantum refrigerators.
    While localized thermalization does not affect the refrigerator's cooling capabilities, nonlocal thermalization markedly alters the refrigerator's equilibrium properties, when constrained to separable dynamics.
    For the correlated dephasing of a two-qubit system, we demonstrated that entanglement prevents heat exchange with its surroundings. 
    Finally, we considered a qubit coupled to a finite-dimensional environment to gauge the impact of system-environment entanglement on the exchange of heat.
    We found that entanglement limits the total amount of heat that is generated in the process.

    The thermodynamics of constrained quantum systems may be a fruitful endeavor for future research.
    Beyond the fundamental aspects studied in this paper, it may be interesting to investigate additional thermodynamic performance indicators of constrained systems, such as the Carnot optimal power~\cite{CPA13}.
    This provides an operational measure for the cooling capabilities of a quantum refrigerator and for the work extraction of a quantum engine. 
    A more ambitious project could even realize the constrained dynamics experimentally by performing continuous measurements on a compound system.
    For this, the constrained dynamics of the probe must be calculated beforehand to devise the sequence of measurements that project a trajectory onto the closest product state.
    Moreover, the separability Lindblad equation, governing the constrained dynamics, poses several theoretical challenges.
    It is a nonlinear equation of motion with a generically time-dependent Lindbladian, and thus its steady states are neither unique nor do they generally exist.
    Finding convenient ways to calculate the steady state of a constrained evolution would significantly simplify the study of the equilibrium thermodynamics when a full-time propagation of the constrained dynamics can be avoided. 

    The notion of a constrained system is ubiquitous in physics \cite{AB51}, and its applications range from differential geometry \cite{A10} and quantum information \cite{PM24} to nonlinear programming \cite{AGL19}.
    While having its origin in Lagrangian mechanics, it applies to any dynamical (quantum) system.
    This work utilized such constraints as a path to characterize the role that entanglement plays in quantum thermodynamics.
    Unlike existing approaches, we do not simply compute a physical property of a system in parallel with an entanglement measure, signifying a mere correlation between the property and the presence of entanglement.
    We believe that imposing a constraint on a system more closely resembles the effect of a (surgical) intervention in which we remove a quantum correlation by hand \cite{P09}---answering how much of a system's change is \textit{caused by} (not correlated with) its quantumness.

    \acknowledgements
    We gratefully acknowledge financial support from Danmarks Grundforskningsfond (DNRF 139, Hy-Q Center for Hybrid Quantum Networks). J.A. acknowledges support from the Novo Nordisk Foundation (Challenge project “Solid-Q”), and J.P. from the Alexander von Humboldt Foundation (Feodor Lynen Research Fellowship).
    L.A. and J.S. acknowledge the financial support through the QuantERA project QuCABOoSE.

    \appendix 

    \section{Algorithmic Monte Carlo wavefunction approach for dynamics constrained to separable states}
    \label{app:numerics}

    Here, we recap the Monte Carlo wavefunction method for simulating the dynamics of an open quantum system constrained to separable states \cite{AP24}. 
    We consider a Markovian dissipation process governed by the quantum master equation~\eqref{eq:LMEq} with Hamiltonian $H$ and jump operators $L_k$.
    Note that the master equation~\eqref{eq:LMEq} is invariant under the inhomogeneous transformation
    \begin{equation}
        \label{eq:shifted-ops}
        \begin{split}
            H & \mapsto H + \frac{1}{2i}\sum_k\left(\lambda_{k}^*L_k-\lambda_{k}L_k^\dagger \right),\\
            L_k & \mapsto L_k + \lambda_{k}\mathbb{1},\\
        \end{split}
    \end{equation}
    where $\lambda^{k}$ is a complex number.
    We choose $\lambda^{k}\gg\norm{L_k}$ with $\norm{\cdot}$ being a matrix norm of one's choice. 
    This ensures that the jump operator become small perturbations of the identity, $L_k + \lambda_{k}\mathbb{1}$, and thus the state $L_k\ket{\psi}$ is only slightly altered before being projected onto the closest product state; see Ref.~\cite{AP24} for a complete derivation.
    Throughout, we assume the operators $H$ and $L_k$ have undergone this transformation.

    From the separability Lindblad equation~\eqref{eq:SepLin}, we identify a non-Hermitian Hamiltonian, 
    \begin{equation}
        \label{eq:non-Herm-constrained}
        \begin{split}
            \hat{H}_{\ms} &= \frac{i(n-1)}{2}\sum_k\langle L_k^\dagger L_k\rangle \mathbb{1}^{\otimes n}\\
            &\quad +\sum_{d=1}^{n}\Big[\mathbb{1}^{\otimes d-1}\otimes(H)_d\otimes\mathbb{1}^{\otimes n-d} \\
            &\quad -\frac{i}{2}\sum_k\mathbb{1}^{\otimes d-1}\otimes(L_k^\dagger L_k)_d\otimes\mathbb{1}^{\otimes n-d}\Big],
        \end{split}
    \end{equation}
    which governs a small time step $\tau$ of the constrained evolution, 
    \begin{equation}
        \ket{\psi(t+\tau)} \propto e^{-i \hat{H}_{\ms}(t) \tau} \ket{\psi(t)},
    \end{equation}
    of the product state $\ket{\psi}=\bigotimes_{d=1}^{n}\ket{\psi_d}$ of an $n$-partite system. 
    The deterministic evolution is interrupted by occasional 
    quantum jumps,
    \begin{equation}
    \ket{\psi}\mapsto\frac{\bigotimes_{d=1}^{n}(L_k)_d}{\langle L_k \rangle^{(n-1)}}\ket{\psi},
    \end{equation}    
    which are constrained to product states.
    Note that $\hat{H}_{\ms}$ and $\bigotimes_{d=1}^{n}(L_k)_d$ depend explicitly on the state $\ket{\psi(t)}$ via Eq.~\eqref{eq:sep-op}. 
    
    The algorithm realizing the constrained Monte Carlo wavefunction method consists of the following steps:
    \begin{enumerate}
        \item At time $t_m\geq 0$, pick a random number $r\in(0,1)$.
        \item Compute $\mathcal{N}=\langle{\psi(t_m)}\ket{\psi(t_m)}$.
        \item If $\mathcal{N}>r$, the system evolves deterministically: 
        
        \begin{enumerate}[leftmargin=0pt]
            \item[3.1] Compute $\hat{H}_{\ms}(t_m)$ for $\ket{\psi(t_m)}=\bigotimes_{d=1}^{n}\ket{\psi_d(t_m)}$ using Eq.~\eqref{eq:non-Herm-constrained} and evolve the state for time $\tau/2$,
            \begin{equation*}
            \centering
            \ket{\psi(t_m+\tau/2)}=e^{-i\tau \hat{H}_{\ms}(t_m)/2}\ket{\psi(t_m)}.
            \end{equation*}
            \item[3.2] Compute $\hat{H}_{\ms}(t_m+\tau/2)$ for $\ket{\psi(t_m+\tau/2)}$ using Eq.~\eqref{eq:non-Herm-constrained}.
            \item[3.3] Evolve $\ket{\psi(t_m)}$ for time $\tau$ using $\hat{H}_{\ms}(t_m+\tau/2)$,
            \begin{equation*}
            \centering
            \ket{\psi(t_m+\tau)}=e^{-i\tau \hat{H}_{\ms}(t_m+\tau/2)}\ket{\psi(t_m)}.
            \end{equation*}
        \end{enumerate} 
        
        \item If $\mathcal{N}<r$, a quantum jump occurs: 

        \begin{enumerate}[leftmargin=0pt]
            \item[4.1] Pick $k$ from the distribution $p_k=\Tilde{p}_k/\sum_k \Tilde{p}_k$ where
            \begin{equation*}
            \Tilde{p}_k=\frac{1}{\abs{\langle L_k \rangle}^{2(n-1)}}\bra{\psi(t_m)}\bigotimes_{d=1}^{n}(L_k^\dagger)_d(L_k)_d\ket{\psi(t_m)}.
            \end{equation*}
            \item[4.2] Given $k$, the quantum jump is realized as
            \begin{equation*}
             \ket{\psi(t_m+\tau)} = \frac{\bigotimes_{d=1}^{n}(L_k)_d}{\langle L_k \rangle^{(n-1)}}\ket{\psi(t_m)}.
         \end{equation*}
        \end{enumerate}
        
        \item From the normalized product state,
        \begin{equation*}
            \rho_\ms (t_m + \tau) = \frac{\ket{\psi(t_m+\tau)}\bra{\psi(t_m+\tau)}}{\langle\psi(t_m+\tau)\ket{\psi(t_m+\tau)}},
        \end{equation*}
        one computes the mean of an observable $F$ as 
        \begin{equation*}
            \braket{F(t_m + \tau)}_\ms =  \Tr[F\rho_\ms(t_m + \tau)].
        \end{equation*}
        \item Repeat 2. to 5. until the last time step, $t_m = t_{\mathrm{end}}$.
        \item Repeat 1. to 6. for $J$ trajectories and take the average, $\overline{\braket{F(t)}_\ms} = \tfrac{1}{J}\sum_{j=1}^J \braket{F(t)}_{\ms,j}$ for $t\in[0,t_{\mathrm{end}}]$. 
    \end{enumerate}

    For details on numerical schemes for solving the nonlinear Schrödinger equation governing the deterministic part of the above algorithm see Ref.~\cite{OWA26}. 
    
    \section{Jump operators for the delocalized model}
    \label{app:delocalized}

    Here, we list the jump operators, $L_{k,\omega}=\Gamma_{k,\omega} A_{k,\omega}$, for the delocalized model of the three-qubit quantum refrigerator in Sec.~\ref{sec:fridge}. 
    The damping rates are
    \begin{equation}
        \Gamma_{k,\omega} = \gamma\omega^3e^{\beta_k\omega/2}\sinh(\beta_k\omega/2),
    \end{equation}
    with $\beta_k=1/T_k$ being the inverse temperature of the $k$th bath, and the operators $A_{k,\omega}$ are given by \cite{CPA13}
    \begin{equation*}
        \begin{split}
            A_{\mathrm{w},\omega_\mathrm{w}}&=\ket{1}\bra{2} + \ket{6}\bra{3},\\ 
            A_{\mathrm{w},\omega_\mathrm{w} + g}&=(\ket{4}\bra{8} - \ket{7}\bra{5})/\sqrt{2},\\ 
            A_{\mathrm{w},\omega_\mathrm{w} - g}&=(\ket{4}\bra{7} + \ket{8}\bra{5})/\sqrt{2},\\
            A_{\mathrm{h},\omega_\mathrm{h}}&=\ket{2}\bra{5} + \ket{4}\bra{6},\\ 
            A_{\mathrm{h},\omega_\mathrm{h} + g}&=(\ket{7}\bra{3} + \ket{1}\bra{8})/\sqrt{2},\\ 
            A_{\mathrm{h},\omega_\mathrm{h} - g}&=(\ket{8}\bra{3} - \ket{1}\bra{7})/\sqrt{2},\\
            A_{\mathrm{c},\omega_\mathrm{c}}&=\ket{1}\bra{4} + \ket{5}\bra{3},\\ 
            A_{\mathrm{c},\omega_\mathrm{c} + g}&=(\ket{2}\bra{8} - \ket{7}\bra{6})/\sqrt{2},\\ 
            A_{\mathrm{c},\omega_\mathrm{c} - g}&=(\ket{2}\bra{7} + \ket{8}\bra{6})/\sqrt{2},\\
        \end{split}
    \end{equation*}
    where we defined
    \begin{equation*}
        \begin{split}
            \ket{1}&=\ket{000},\quad \ket{2}=\ket{100},\quad \ket{3}=\ket{111},\\
            \ket{4}&=\ket{001},\quad \ket{5}=\ket{110},\quad \ket{6}=\ket{011},\\
            \ket{7}&=(\ket{101}-\ket{010})/\sqrt{2},\quad\,~ \ket{8}=(\ket{101}+\ket{010})/\sqrt{2},\\
        \end{split}
    \end{equation*}  
    and used the shorthand $\ket{jkl}=\ket{j_\mathrm{w}}\otimes\ket{k_\mathrm{h}}\otimes\ket{l_\mathrm{c}}$. 
    Additionally, the master equation for this model contains the jump operators $L_{k,\omega}^\prime=\Gamma_{k,-\omega} A_{k,\omega}^\dag$.
    
    \section{Exchange interaction}
    \label{app:exchange}

    Here, we derive both the free and constrained evolution of two qubits $A$ and $B$ interacting according to
    \begin{equation}
        \label{eq:A1}
        H = H_A\otimes\mathbb{1} + \mathbb{1}\otimes H_B + \kappa V,
    \end{equation}
    where $H_{A}=\omega_A\ket{1}\bra{1}$ and $H_{B}=\omega_B\ket{1}\bra{1}$ are the free Hamiltonians and $V$ is the swap operator applied with strength $\kappa$.
    The system is initially in the state
    \begin{equation}
        \label{eq:A2}
        \begin{split}
            \rho_0 & = \ket{\psi_A}\bra{\psi_A} \otimes \rho_B,\\
        \end{split}
    \end{equation}
    where 
    \begin{equation}
        \ket{\psi_A}=\frac{1}{\sqrt{Z_A}}\big(\ket{0}+ e^{-\beta_A \omega_A/2}\ket{1}\big)
    \end{equation}
    is a coherent superposition and $\rho_B = e^{-\beta_B H_B}/Z_B$ is a thermal state with $Z_B = 1 + e^{-\beta_B \omega_B}$.

    \subsection{Time evolution and heat exchange}

    To solve the equations of motion for the Hamiltonian~\eqref{eq:A1}, we transform the problem into the interaction picture.
    The joint state at time $t$ is given by
    \begin{equation*}
        \hat{\rho}(t)=\cos^2(\kappa t)\rho_0 + i\cos(\kappa t)\sin(\kappa t)[\rho_0,V] + \sin^2(\kappa t)V \rho_0 V,
    \end{equation*}
    with $\rho_0=\hat{\rho}(0)$ given in Eq.~\eqref{eq:A2} and
    \begin{equation}
        \label{eq:int-pic}
         \hat{\rho}(t)=e^{i H_A t}\otimes e^{i H_B t} \rho(t)e^{-i H_A t}\otimes e^{-i H_B t}    
    \end{equation}
    is the interaction picture density matrix. 
    Tracing out $B$ yields the reduced state of $A$,
    \begin{equation}
        \label{eq:A3}
        \begin{split}
            \rho_A(t)&=\Tr_B[\rho(t)]\\
            &=e^{-i H_A t}\big[\cos^2(\kappa t) \ket{\psi_A}\bra{\psi_A}\\
            &\quad + i\cos(\kappa t)\sin(\kappa t)\sigma + \sin^2(\kappa t)\rho_B\big]e^{i H_A t},\\
        \end{split}
    \end{equation}
    where we defined
    \begin{equation*}
        \begin{split}
            \sigma & = \Tr_B([\rho_0,V])\\
            & = \frac{1}{Z_B}\Tr_B(\ket{\psi_A 0}\bra{0 \psi_A} + e^{-\beta_B \omega_B}\ket{\psi_A 1}\bra{1 \psi_A}) - \mathrm{H.c.}\\
            & = \frac{1}{Z_B}(\ket{\psi_A}\bra{\phi} - \ket{\phi}\bra{\psi_A}),
        \end{split}
    \end{equation*}
    with a vector
    \begin{equation}
        \ket{\phi}=\frac{1}{\sqrt{Z_A}}\big(\ket{0} + e^{-\beta_A \omega_A/2-\beta_B\omega_B}\ket{1}\big).
    \end{equation}
    The heat exchange is found to be
    \begin{equation}
        \begin{split}
            Q(t) & = \Tr\{H_A[\rho_A(t) - \ket{\psi_A}\bra{\psi_A}]\}\\
            &=[\cos^2(\kappa t) - 1]\braket{H_A}_{\psi_A} + \sin^2(\kappa t)\braket{H_A}_{\rho_B},\\
        \end{split}
    \end{equation}
    where we used Eq.~\eqref{eq:A3} and the fact that $\sigma$ obeys $\braket{H_A}_{\sigma}=\braket{\phi|H_A|\psi_A}/Z_B - \mathrm{c.c.}=0$.
    Inserting the mean energies of the initial state~\eqref{eq:A2},
    \begin{equation*}
        \begin{split}
            \braket{H_A}_{\psi_A}&=\frac{\omega_A e^{-\beta_A\omega_A}}{1 + e^{-\beta_A\omega_A}},\quad \braket{H_A}_{\rho_B}=\frac{\omega_A e^{-\beta_B \omega_B}}{1 + e^{-\beta_B \omega_B}},
        \end{split}
    \end{equation*}
    yields Eq.~\eqref{eq:phys-heat-TLS}.

    \subsection{Constrained time evolution and heat exchange}

    We solve the constrained equations of motion in the interaction picture~\eqref{eq:int-pic}.
    The initial state,
    \begin{equation}
        \label{eq:init-state-exp}
        \rho_0=\frac{1}{Z_B}\big(\ket{\psi_A0}\bra{\psi_A0} + e^{-\beta_B \omega_B} \ket{\psi_A1}\bra{\psi_A1}\big),
    \end{equation}
    corresponds to a mixture of two pure-state trajectories.
    These have to be propagated individually because Eq.~\eqref{eq:SSE} is a nonlinear differential equation \cite{SW20}.
    The operator for subsystem $A$ acts as 
    \begin{equation}
        \begin{split}
            (V)_A\ket{\psi_A(t)} & = \bra{\psi_B(t)}V\ket{\psi_A(t)\psi_B(t)}\\
            & = \braket{\psi_B(t)\vert\psi_A(t)} \ket{\psi_B(t)},\\
        \end{split}
    \end{equation}
    which implies $(V)_A=\ket{\psi_B(t)}\bra{\psi_B(t)}$.
    Similarly, we find $(V)_B=\ket{\psi_A(t)}\bra{\psi_A(t)}$. 
    The separable equations of motion~\eqref{eq:SSE} are solved by \cite{SW17}
    \begin{equation}
        \label{eq:A4}
        \begin{split}
            \ket{\psi_A(t)} & = \cos(|q|\kappa t)\ket{\psi_A} - i\frac{q^*}{|q|}\sin(|q|\kappa t)\ket{\psi_B},\\
            \ket{\psi_B(t)} & = \cos(|q|\kappa t)\ket{\psi_B} - i\frac{q}{|q|}\sin(|q|\kappa t)\ket{\psi_A},\\
        \end{split}
    \end{equation}
    where $q=\braket{\psi_A|\psi_B}$ denotes the overlap between arbitrary initial states $\ket{\psi_A}$ and $\ket{\psi_B}$.
    For the two initial configurations $\ket{\psi_A}\otimes \ket{j}$, with $j\in\{0,1\}$, we get
    \begin{equation}
        \label{eq:A5}
        \begin{split}
            \ket{\psi_{j}(t)}&=\big[\cos(q_j\kappa t)\ket{\psi_A} - i\sin(q_j\kappa t) \ket{j}\big]\\
            &\quad \otimes \big[\cos(q_j\kappa t)\ket{j} - i\sin(q_j\kappa t) \ket{\psi_A}\big],\\
        \end{split}
    \end{equation}
    with overlaps 
    \begin{equation*}
        \begin{split}
            q_0 & = \braket{0|\psi_A}=\frac{1}{\sqrt{Z_A}},\quad
            q_1 = \braket{1|\psi_A}=\frac{e^{-\beta_A\omega_A/2}}{\sqrt{Z_A}}.
        \end{split}
    \end{equation*}
    From Eq.~\eqref{eq:A5}, we obtain the reduced state of $A$ as
    \begin{equation}
        \label{eq:red-traj}
        \ket{\psi_{A,j}(t)} = \cos(q_j\kappa t)\ket{\psi_A} - i\sin(q_j\kappa t) \ket{j}.
    \end{equation}
    The constrained state of qubit $A$ is obtained by weighting the trajectories~\eqref{eq:red-traj} with probabilities specified in Eq.~\eqref{eq:init-state-exp}, leading to
    \begin{equation*}
        \rho_{\ms}(t) = \frac{1}{Z_B}\big(\ket{\psi_{A,0}}\bra{\psi_{A,0}} + e^{-\beta_B \omega_B}\ket{\psi_{A,1}}\bra{\psi_{A,1}}\big).
    \end{equation*}
    The heat exchange through separable states becomes
    \begin{equation*}
        \begin{split}
            Q_{\ms}(t) & = \Tr\{H_A[\rho_{\ms}(t)-\ket{\psi_A}\bra{\psi_A}]\}\\
            & = \frac{1}{Z_B}\left(\braket{H_A}_{\psi_{A,0}} + e^{-\beta_B \omega_B}\braket{H_A}_{\psi_{A,1}}\right) - \braket{H_A}_{\psi_A}.
        \end{split}
    \end{equation*}
    Calculating the mean energies of each trajectory
    \begin{equation}
        \begin{split}
            \braket{H_A}_{\psi_{A,0}} & = \braket{H_A}_{\psi_A}\cos^2\big(\tfrac{\kappa t}{Z_A}\big),\\
            \braket{H_A}_{\psi_{A,1}} & = \omega_A \sin^2(q_1\kappa t) + \braket{H_A}_{\psi_A}\cos^2(q_1\kappa t),\\
        \end{split}
    \end{equation}
    provides an explicit expression for the constrained heat,
    \begin{equation*}
        \begin{split}
            Q_{\ms}(t) & = \frac{1}{Z_B}\big[\cos^2(q_0\kappa t) + e^{-\beta_B \omega_B}\cos^2(q_1\kappa t)\big]\braket{H_A}_{\psi_A}\\
            & \quad + \omega_Ae^{-\beta_B \omega_B} \sin^2(q_1\kappa t) - \braket{H_A}_{\psi_A}.
        \end{split}
    \end{equation*}
    Inserting the explicit form of $\braket{H_A}_{\psi_A}$ into $Q_\ms$ yields Eq.~\eqref{eq:sep-heat-TLS}.

\end{document}